\documentstyle[epsfig]{mn}

\newif\ifAMStwofonts

\def\gtsima{$\; \buildrel > \over \sim \;$}
\def\ltsima{$\; \buildrel < \over \sim \;$}
\def\gsim{\lower.5ex\hbox{\gtsima}}
\def\lsim{\lower.5ex\hbox{\ltsima}}
\newcommand{\etal}{et al.\ }
\def\Msun{{M_\odot}}
\def\Zsun{{Z_\odot}}
\def\be{\begin{equation}}
\def\ee{\end{equation}}
\def\ie{{\frenchspacing\it i.e. }}
\ifoldfss
  \ifCUPmtlplainloaded \else
    \NewTextAlphabet{textbfit} {cmbxti10} {}
    \NewTextAlphabet{textbfss} {cmssbx10} {}
    \NewMathAlphabet{mathbfit} {cmbxti10} {} % for math mode
    \NewMathAlphabet{mathbfss} {cmssbx10} {} %  "   "    "
  \fi
  \ifAMStwofonts
    \ifCUPmtlplainloaded \else
      \NewSymbolFont{upmath} {eurm10}
      \NewSymbolFont{AMSa} {msam10}
      \NewMathSymbol{\upi}     {0}{upmath}{19}
      \NewMathSymbol{\umu}     {0}{upmath}{16}
      \NewMathSymbol{\upartial}{0}{upmath}{40}
      \NewMathSymbol{\leqslant}{3}{AMSa}{36}
      \NewMathSymbol{\geqslant}{3}{AMSa}{3E}

      \let\leq=\leqslant \let\le=\leqslant
      \let\geq=\geqslant \let\ge=\geqslant
    \fi
  \fi
\fi % End of OFSS

\ifnfssone
  \newmathalphabet{\mathit}
  \addtoversion{normal}{\mathit}{cmr}{m}{it}
  \addtoversion{bold}{\mathit}{cmr}{bx}{it}
  \newmathalphabet{\mathbfit} % math mode version of \textbfit{..}
  \addtoversion{normal}{\mathbfit}{cmr}{bx}{it}
  \addtoversion{bold}{\mathbfit}{cmr}{bx}{it}
  \newmathalphabet{\mathbfss} % math mode version of \textbfss{..}
  \addtoversion{normal}{\mathbfss}{cmss}{bx}{n}
  \addtoversion{bold}{\mathbfss}{cmss}{bx}{n}
  \ifAMStwofonts
    \ifCUPmtlplainloaded \else
      \UseAMStwoboldmath
      \makeatletter
      \new@mathgroup\upmath@group
      \define@mathgroup\mv@normal\upmath@group{eur}{m}{n}
      \define@mathgroup\mv@bold\upmath@group{eur}{b}{n}
      \edef\UPM{\hexnumber\upmath@group}
      \new@mathgroup\amsa@group
      \define@mathgroup\mv@normal\amsa@group{msa}{m}{n}
      \define@mathgroup\mv@bold\amsa@group{msa}{m}{n}
      \edef\AMSa{\hexnumber\amsa@group}
      \makeatother
      \mathchardef\upi="0\UPM19
      \mathchardef\umu="0\UPM16
      \mathchardef\upartial="0\UPM40
      \mathchardef\leqslant="3\AMSa36
      \mathchardef\geqslant="3\AMSa3E

      \let\leq=\leqslant \let\le=\leqslant
      \let\geq=\geqslant \let\ge=\geqslant
    \fi
  \fi
\fi % End of NFSS release 1

\ifnfsstwo
  \DeclareMathAlphabet{\mathbfit}{OT1}{cmr}{bx}{it}
  \SetMathAlphabet\mathbfit{bold}{OT1}{cmr}{bx}{it}
  \DeclareMathAlphabet{\mathbfss}{OT1}{cmss}{bx}{n}
  \SetMathAlphabet\mathbfss{bold}{OT1}{cmss}{bx}{n}
  \ifAMStwofonts
    \ifCUPmtlplainloaded \else
      \DeclareSymbolFont{UPM}{U}{eur}{m}{n}
      \SetSymbolFont{UPM}{bold}{U}{eur}{b}{n}
      \DeclareSymbolFont{AMSa}{U}{msa}{m}{n}
      \DeclareMathSymbol{\upi}{0}{UPM}{"19}
      \DeclareMathSymbol{\umu}{0}{UPM}{"16}
      \DeclareMathSymbol{\upartial}{0}{UPM}{"40}
      \DeclareMathSymbol{\leqslant}{3}{AMSa}{"36}
      \DeclareMathSymbol{\geqslant}{3}{AMSa}{"3E}

      \let\leq=\leqslant \let\le=\leqslant
      \let\geq=\geqslant \let\ge=\geqslant
    \fi
  \fi
\fi % End of NFSS release 2

\ifCUPmtlplainloaded \else
  \ifAMStwofonts \else % If no AMS fonts
    \def\upi{\pi}
    \def\umu{\mu}
    \def\upartial{\partial}
  \fi
\fi

\title{Cosmic stellar relics in the Galactic halo}
\author[Stefania Salvadori, Raffaella Schneider \& Andrea Ferrara]
{Stefania Salvadori$^{1}$, Raffaella Schneider$^{2}$  \& Andrea Ferrara$^{1}$  \\
$^1$SISSA/International School for Advanced Studies, Via Beirut 4, 34100 Trieste, Italy\\ 
$^2$INAF/Osservatorio Astrofisico di Arcetri, Largo Enrico Fermi 5, 50125 Firenze, Italy}
\date{}

\pagerange{\pageref{firstpage}--\pageref{lastpage}}
\pubyear{}

\begin{document}

\maketitle 

\label{firstpage}

\begin{abstract}
We study the stellar population history and chemical evolution
of the Milky Way in a hierarchical $\Lambda$CDM model for
structure formation. Using a Monte Carlo method based on the
semi-analytical extended Press \& Schechter formalism, we develop a
new code GAMETE (GAlaxy MErger Tree \& Evolution) to reconstruct 
the merger tree of the Galaxy and follow the evolution of gas and stars
along the hierarchical tree. Our approach allows us to compare the 
observational properties of the Milky Way with model results, exploring
different properties of primordial stars, such as their Initial Mass
Function (IMF) and the critical metallicity for low-mass star formation, 
$Z_{\rm cr}$. In particular, by matching our predictions to the metallicity 
distribution function of metal-poor stars in the Galactic halo we find that:
(i) a strong supernova feedback is required to reproduce the observed properties of the Milky Way; 
(ii) stars with [Fe/H]$<-2.5$ form in halos accreting Galactic Medium (GM) enriched by earlier supernova explosions;
(iii) the fiducial model ($Z_{cr} = 10^{-4} Z_\odot$, $m_{PopIII}=200\Msun$) provides an overall good fit to the MDF, but cannot
  account for the two HMP stars with [Fe/H]$<-5$; the latter can be accommodated if $Z_{cr}\le 10^{-6}\Zsun$ but 
  such model overpopulates the ``metallicity desert", \ie the range $-5.3<$ [Fe/H] $<-4$ in which no
  stars have been detected;
(iv) the current non-detection of metal-free stars robustly constrains either $Z_{cr}>0$ or the masses of the first stars
  $m_{PopIII}>0.9\Msun$;
(v) the statistical impact of truly second generation stars, \ie stars forming out of gas polluted {\it only} by
  metal-free stars, is negligible in current samples; 
(vi) independently of $Z_{cr}$,  $60\%$ of metals in the Galactic Medium are ejected through winds by halos with
 masses $M<6\times 10^9\Msun$, thus showing that low-mass halos are the dominant population contributing to
 cosmic metal enrichment. We discuss the limitations of our study and comparison with previous work.
\end{abstract}


\section{Introduction}

The first stars represent the first source of light, metals and dust after
the Big Bang. Since they form out of gas of primordial composition, 
they are expected to be metal free. For this reason, they have
been historically called population III stars (Pop III) to
make a distinction from the observed population II/I stars, 
which have typical metallicities $Z \gsim 10^{-5}-10^{-4}$ and
$Z\sim 0.02$, respectively (Bromm \& Larson 2004). 
The impact of first stars on the subsequent structure formation
history is very strong through different feedback mechanisms: 
UV photons from the first stars are likely responsible for the
reionization and heating of the intergalactic 
medium (IGM); heavy elements are synthesized inside such stars and during the 
early supernova (SN) explosions dust grains are formed. Dust and metals 
are released in the interstellar medium (ISM) and eventually 
enrich the IGM through SN winds. The efficiency of all
these processes depends on the properties of the first stars, which
therefore control the evolution of the ISM, IGM and 
of the cosmic environment in which galaxies form. 

There are still many puzzling questions about the first stars, 
related to the lack of observational evidences of such pristine objects. 
At present, the primordial Initial Mass Function (IMF) 
represents one of the most discussed topics in this field. 
Theoretical studies support the idea of a top-heavy IMF, biased towards massive
and very massive stars (Omukai \& Nishi 1998; Abel, Bryan \& Norman 2002; 
Bromm, Coppi \& Larson 2002; Ripamonti \etal 2002; Nakamura \& Umemura 2002; 
Schneider \etal 2002; Omukai \& Palla 2003; Bromm \& Loeb 2004;
O'Shea \& Norman 2006; Yoshida \etal 2006). Indeed, in the absence of
metals, the collapsing star forming gas is predicted to fragment 
in $\sim 10^3 \Msun$ clumps, progenitors of the stars which will later 
form in their interiors accreting gas on the central proto-stellar core. 
Because of the poor understanding of the complex physics which controls 
the late phases of accretion and protostellar feedback, the final stellar 
masses are still largely uncertain but likely to be in the range 
$30 \Msun - 300 \Msun$ (Tan \& McKee 2004).

In contrast with the above speculations, present-day star formation 
(Pop II/I stars) is known to be characterized by a Salpeter IMF in the mass 
range $0.1 \Msun - 100 \Msun$, flattening below masses of $0.35M_{\odot}$ (Larson 1998).
Since the characteristic mass of local Pop~II/I stars is $\sim 1 \Msun$, 
if our current understanding of primordial star formation is correct, there
must have been a transition in the properties of star-forming 
regions through cosmic times. Recent theoretical studies suggest that 
the initial metallicity of the star-forming gas represents the key element 
controlling this transition (Bromm \etal 2001; Omukai \etal 2001, 2005; 
Schneider \etal 2002, 2003, 2006; Bromm \& Loeb 2004). 
Following the evolution of protostellar gas clouds with different 
values of the initial metallicity and including dust and molecules 
as cooling agents, Schneider \etal (2002-2006) and Omukai \etal (2005) 
find that when the metallicity is in the critical range
$10^{-6} < Z_{cr}/Z_\odot < 10^{-4}$ there occurs a transition
in fragmentation scales from $\sim 10^3\Msun$ to solar or subsolar fragments
and argue that this is an indication for a transition in 
characteristic stellar masses. The critical metallicity value depends on the
fraction of metals locked in dust grains, which provide an additional
cooling channel at high densities enabling fragmentation to solar and subsolar
clumps (Schneider \etal 2003, 2006; Omukai \etal 2005; Tsuribe \& Omukai 2006). 
Thus, according to this scenario the onset of low-mass star formation in the 
Universe is triggered by the presence of metals and dust in the parent clouds
to levels exceeding $Z_{cr}$. 

In this scenario, stellar archeology of the most metal-poor stars represents 
a promising way to explore primordial star formation. Due to their low
metallicities, metal-poor stars are expected to form out of gas enriched 
by members of the first stellar generation. Therefore, despite of the fact 
that metal-poor stars are observed in the Galactic halo or in nearby 
dwarf galaxies, satellites of the Milky Way (MW), these objects might well be
living fossils of the first star formation episodes in the Universe.
One of the most important observational constraints is provided by the
distribution of stellar metallicities in the halo of the Galaxy, 
the so-called Metallicity Distribution Function (MDF). In their 
early studies, Ryan \& Norris (1991) and Carney \etal (1996) show
that this function peaks at [Fe/H]$ =-1.6$\footnote{Hereafter 
we use the notation [Fe/H]$=\log(N_{Fe}/N_{H}) - \log(N_{Fe}/N_{H})_{\odot}$ 
normalized to the solar values of Anders \& Grevesse (1989).} 
with a tail extending down to [Fe/H] $\approx -4$. Although the compilation of 
Ryan \& Norris (1991) is relatively old and consists only of 300 stars, 
their MDF is consistent with that obtained by more recent and 
larger samples collected by the HK survey of Beers \etal (1992) and 
by the Hamburg/ESO objective prism survey (HES, Wisotzki \etal 2000; 
Christlieb 2003)
whose analysis is still underway to quantify completeness and selection effects 
(see Beers \& Christlieb 2005 for a thorough review). The joint HK and HES sample
includes $2756$ stars with [Fe/H]$<-2$. It will be further extended by the
large number of metal-poor stars contained in the most recent public data release
of the Sloan Digital Sky Survey (York \etal 2000), and its planned extension which
includes the program SEGUE, specifically targeted to collect metal-poor stars with
[Fe/H]$<-2$ and to constrain the transition from the disk population to the halo.   

The HES has recently revealed the existence of two stars 
with [Fe/H]$=-5.3$\footnote{A recent analysis based on the detection
  of Fe~II lines leads to a downward revision of the iron abundance of
  HE0107-5240 to [Fe/H] = $-5.7 \pm 0.2$ (Christlieb, Bessell \&
  Eriksson 2006).} (HE0107-5240, Christlieb \etal 2002) and
[Fe/H]$=-5.4$ (HE1327-2326, Frebel \etal 2005). Both of these stars  
show a large overabundance of carbon and nitrogen with respect to
iron and other heavy elements, suggesting a similar origin for their
abundance pattern. Essentially two classes of models have been proposed to interpret the
observed abundances: (i) the stars were born in a primordial environment and
had their surfaces polluted at a later time  
(Shigeyama, Tsujimoto \& Yoshii 2003; Suda \etal 2004); (ii) they
had been pre-enriched by the yields of one or more SNe (Christlieb
\etal 2002, 2004; Umeda \& Nomoto 2003; Bonifacio, Limongi \& Chieffi 2003;
Schneider \etal 2003; Iwamoto \etal 2005). In the first case, these two
stars would be low-mass Pop III stars, in conflict with the proposed 
critical metallicity scenario. In the second case, instead, these stars
would be second generation stars enriched by the ashes of the first 
SNe. At present, the available evidence does not allow to exclude 
any of the above proposed scenarios but shows that the observational data
on metal poor stars are starting to challenge theoretical models.   

This challenge is even more serious as it resonates with 
additional constraints on the primordial IMF placed by completely
independent observations. In particular, the recent downward 
revision of Thomson scattering optical depth found in the 
analysis of the {\it Wilkinson Microwave Anisotropy Probe} (WMAP) 
three years data (Page \etal 2006) predicts a later reionization epoch, 
in agreement with a universal Salpeter IMF (Choudhury \& Ferrara 2006; 
Gnedin \& Fan 2006). The same conclusion is drawn when theoretical models 
(Salvaterra \etal 2006) are confronted with the observed anisotropies 
in the near-infrared background (Kashlinsky \etal 2005). 
These results tend to favor a relatively unevolving Salpeter IMF up to
the highest redshifts. 

The motivation of the present study was to explore to what extent the
shape of the low-metallicity tail of the halo MDF can provide insights 
on the nature of the first stars. In particular, we aim at addressing
the following questions: \\
$\bullet$ What are the implications of the observed low-metallicity tail of
the halo MDF?\\
$\bullet$ Does a sharp MDF cutoff put constraints on the primordial IMF? \\
$\bullet$ What is the relation between the MDF cutoff and the critical metallicity ? \\
$\bullet$ What is the origin of the ``metallicity desert", \ie the absence of stars in 
the range $-5.3<$ [Fe/H] $<-4$? \\
$\bullet$ What is the statistical impact in the observed samples of truly 
second generation stars, \ie stars enriched only by the ashes of the
first stellar explosion, which then can be used to constrain the nucleosynthetic
products and the masses of primordial stars?

Several authors have attempted to interpret the implications of the observed
halo MDF by applying different methods.  
Using a semi-analytical model, Hernandez \& Ferrara (2001) 
deduced that a characteristic Pop~III mass increasing towards high redshift is required
in order to fit the low-$Z$ tail of the MDF. Prantzos (2003) also pointed out that
the stellar metallicity distribution depends sensitively on whether instantaneous recycling 
is adopted or relaxed. The possible existence of a low-$Z$ cutoff in the MDF has
been interpreted by Oey (2003) in the framework of  stochastic chemical enrichment models, as 
a result of the various metal diffusion/transport/mixing processes at work in the 
Galactic environment. Along similar lines, Karlsson (2006) has noticed that the metallicity  
desert, if confirmed, could be used to
extract information about the past Galactic star formation history.  
Finally, Tumlinson (2006) presented a 
chemical evolution model for the MW which follows
the star formation history in a hierarchical galaxy formation scenario. 
Such model accurately reproduces the halo MDF by Ryan \& Norris (1991). 

The method proposed in this paper is similar in spirit to that of Tumlinson; however,
it introduces a number of novel features, particularly for what concerns the treatment
of mechanical feedback and model calibration.  Moreover, we compare our model results
with a new MDF determination obtained from joint HK and HES data (including $2756$ stars with 
[Fe/H]$\leq -2$) by Beers \& Christlieb (2006).
A detailed comparison between the two studies is given in Sec. 6. 
Using the newly developed code GAMETE (GAlaxy MErger Tree \&
Evolution) we follow the gradual build-up of the stellar 
population and metal enrichment of the MW along its past
hierarchical evolution. The code is based on a Monte Carlo 
algorithm, along the lines of Cole \etal (2000)
and Volonteri, Haardt \& Madau (2003), standing on the Extended Press Schecther (EPS)
theory (Bond \etal 1991; Lacey \& Cole, 1993). We reconstruct the 
hierarchical evolution of the MW from redshift $z=20$ and 
follow, jointly, the chemical enrichment and history of its stellar population. 
The chemical evolution of both proto-Galactic halos and of the  
gas reservoir into which they are embedded, which we call thereafter the {\it Galactic Medium} (GM), 
is computed by including a physically based description of the mechanical
feedback produced by the energy deposition of massive stars. 
In addition, this allows us to evaluate the typical masses of halos which mostly 
contribute to the GM enrichment and to understand how the ejection and accretion of enriched gas
along the hierarchical tree is reflected in the resulting stellar MDF. 
To calibrate the two free parameters of the model (star formation and wind efficiency) 
we compare the inferred Galaxy average properties with their observed values.
The model can then be used to predict the shape of the halo MDF for different 
assumed properties of the first stars, \ie the primordial IMF and critical metallicity.

The paper is organized as follows: in Sec. 2, we describe the
algorithm developed to construct the cosmological\footnote{We adopt a $\Lambda$CDM cosmological model with $h=0.73$, 
$\Omega_{m}=0.24$, $\Omega_{\Lambda} = 0.72$, $\Omega_{b}h^{2}=0.02$, $n=0.95$
and $\sigma_8 = 0.74$, consistent with the 3-yr WMAP data (Spergel \etal 2006).
We adopt the fit to the power spectrum proposed by Bardeen \etal (1986) and modified by Sugiyama (1995).}
merger tree of the Galaxy and we show that it successfully reproduces the 
predictions of the EPS theory. In Sec. 3, we discuss the implementation of 
star formation and mechanical feedback; Sec. 4 discusses the model calibration. 
In Sec. 5 we present the results of the model, which are then compared
with the most recent data in Sec. 6. A brief summary and discussion,
given in Sec. 7, conclude the paper.

\section{Building the merger tree}

The most popular model for galaxy formation ($\Lambda$CDM) 
predicts that galaxies form through a series of  merging processes of
lower mass fragments, in a hierarchical tree picture. We have used a
semi-analytical approach based on the Extended Press \& Schechter
(EPS) theory (Bond \etal 1991; Lacey \& Cole 1993) 
to reconstruct the hierarchical merger history of our Galaxy. 
Halo mass distributions drawn from Press \& Schechter formalism (1974) 
show an excellent agreement with those derived from N-body simulations; its main 
drawback consists in the lack of spatial information.  
There are several alternative methods to construct merger tree algorithms
from EPS  (see Somerville \& Kolatt 1999 for a critical review);  
following Cole \etal (2000) and Volonteri \etal (2003) (herafter VHM), 
we have developed a binary Monte Carlo code with mass accretion.

The main building block of the algorithm follows from equation (2.15) of 
Lacey \& Cole (1993) re-written in terms of the progenitor mass $M$: 
\[
f(M,M_{0})\;dM=\frac{1}{\sqrt{2\pi}}\frac{(\delta_{c}-\delta_{c_{0}})}
{(\sigma_{M}^{2}-\sigma_{M_{0}}^{2})^{3/2}}\;\times
\]
\be
\;\;\;\;\;\;\;\times\;\exp{\left(-\frac{(\delta_{c}-\delta_{c_{0}})^{2}}{2(\sigma_{M}^{2}-\sigma_{M_{0}}^{2})}\right)}\left|\frac{d\sigma_{M}^{2}}{dM}\right|\;dM.
\label{eq:2_1} 
\ee
This equation gives the fraction of mass in a halo of mass $M_{0}$ at
redshift $z_{0}$ which, at an earlier time $z>z_{0}$, belongs to less
massive progenitors having mass in the range $M$ to $M+dM$ (see
also Cole \etal 2000). The quantity $\delta_{c}=\delta_{c}(z)$ 
($\delta_{c_{0}}=\delta_{c}(z_{0})$) is the critical linear overdensity threshold 
for collapse at redshift $z$ ($z_{0}$); $\sigma_{M}^{2}$ ($\sigma_{M_{0}}^{2}$) is  
the linear r.m.s. density fluctuation smoothed with a top-hat filter of mass $M$ ($M_{0}$). 
By construction, the integral of eq.~(\ref{eq:2_1}) from $M=0$ to $M=M_{0}$ must give unity,
\ie all halos were in less massive fragments at earlier epochs.
Multiplying $f(M,M_{0})\;dM$ by $M_{0}/M$, the mass fraction can be translated  into 
the number of halos per unit mass,
\begin{equation}
\frac{dN}{dM}(M,M_{0})\;dM=\frac{M_{0}}{M}\;f(M,M_{0})\;dM.
\label{eq:Nhalo}
\end{equation}
Hereafter we drop the argument $(M,M_{0})$ to simplify the notation.
Taking the limit $z \to z_{0}$ of eq.~(\ref{eq:Nhalo}), one can 
obtain an expression for the average number of progenitors in the mass range $M$, $M+dM$ into which a halo of mass
$M_{0}$ fragments considering a step $dz$ back in time (Cole \etal 2000):

\begin{equation}
\frac{dN}{dM}\;dM=\frac{1}{\sqrt{2\pi}}\frac{1}{(\sigma_{M}^{2}-\sigma_{M_{0}}^{2})^{3/2}}\frac{d\delta_{c}}{dz}\left|\frac{d\sigma_{M}^{2}}{dM}\right|\;dM\;dz,
\label{eq:2_2} 
\end{equation}

\noindent
where $M<M_{0}$ and $z=z_{0}+dz$. This equation represents the core of
our algorithm. We use it to build a binary merger tree that starts from
the present day Galactic dark matter halo and decomposes it into its
progenitors, running backward in time up to redshift $z=20$.
Given the shape of the matter power spectrum in $\Lambda$CDM models, 
the number of halos diverges as the mass $M$ goes to zero.
For this reason, a cutoff (resolution) mass $M_{res}$, 
which marks the transition between {\it progenitors} and {\it mass accretion}, 
is required.
At any given time-step, halos can either (i) lose part of their
mass, corresponding to the cumulative fragmentation into halos with $M < M_{res}$, 
or (ii) fragment into two progenitor halos and lose mass. At each redshift, the mass below
the resolution limit accounts for the GM in which the halos are embedded. 
Note that we refer to {\it mass accretion} as the process of mass loss in the (backward stepping) 
merger algorithm. 
  
Once $M_{res}$ is fixed, one can compute the mean number of progenitors in the mass range $M_{res}<M<M_{0}/2$
that a halo of mass $M_{0}$ at $z_{0}$ is fragmented into during a time-step $dz$

\begin{equation}\label{eq:2_3} 
N_{p}=\int_{M_{res}}^{M_{0}/2}\frac{dN}{dM}\;dM,
\end{equation}

\noindent
and the accreted mass fraction 

\begin{equation}\label{eq:2_4} 
F_{a}=\int_{M_{0}}^{M_{res}}\frac{dN}{dM}\frac{M}{M_{0}}\;dM.
\end{equation}

\noindent
Both these quantities depend on $dz$ (see eq.~\ref{eq:2_2}), suitably chosen to 
prevent multiple fragmentations ($N_{p}<1$). Since the number 
of progenitors decreases with $dz$, binary algorithms generally require 
a high temporal (redshift) resolution.

At each time step and for each progenitor mass $M_0$, we generate a random number $0<R<1$ 
and we compare it with the value $N_p$ derived by eq.~(\ref{eq:2_3}).  If $N_p<R$, the halo does 
not fragment at this step. However a new halo with mass $M_{0}(1-F_{a})$ is produced, 
to account for the accreted matter.
If instead $N_p \geq R$, fragmentation occurs: a new random value in the range $M_{res} < M < M_{0}/2$ is drawn from 
the distribution eq.~(\ref{eq:2_2}) to produce a 
progenitor of mass $M$; mass conservation sets the mass of the second progenitor to $M_{0}(1-F_{a})-M$. 
This procedure, iterated on each progenitor halo and at each redshift, 
provides the whole hierarchical tree of any initially selected galaxy, 
from the present-day up to an initial redshift.

To give a complete picture of the merger tree algorithm 
we need to describe the quantities involved in the equations and specify the adopted free parameters. 
In particular, in eq.~(\ref{eq:2_2}) the expressions 
$\delta_{c}(z)$ and $\delta_{c}(z_{0})$ represent the values 
of the critical density extrapolated to $z=0$. Specifically,
$\delta_{c}(z)=1.686/D(z)$ where $D(z)$ is the linear growth
factor (Carrol, Press \& Turner 1992), 

\begin{equation}\label{eq:2_5} 
D(z)=\frac{5\Omega_{m}(z)}{2(1+z)}\left(\frac{1}{70}+\frac{209}{140}\Omega_{m}(z)-\frac{\Omega^{2}_{m}(z)}{140}+\Omega^{4/7}_{m}(z)\right)^{-1}
\end{equation} 

\noindent
where $\Omega_{m}(z)=\Omega_{m0}(1+z)^{3}[1-\Omega_{m0}+(1+z)^{3}\Omega_{0}]^{-1}$.
For the adopted cosmological parameters the above equation represents a good 
approximation of the linear theory growth factor. On the other hand
the linear rms density fluctuation in eq.~(\ref{eq:2_2}), 
smoothed with a top-hat filter, is given by

\begin{equation}\label{eq:2_6} 
\sigma^{2}(M)=\frac{1}{2\pi}\int_{0}^{\infty}P(k)W^{2}(kR)k^{2}dk,
\end{equation} 

\noindent
where the integral is performed in the $k$ Fourier-space, $P(k)$ is
the cold dark matter power spectrum and 

\[
W(kR)=4\pi R^{3}[(\sin(kR)-kR\cos(kR))/(kR)^{3}],
\]

\noindent
is the top-hat window function. 

We now need to specify the free parameters of the model, $M_{res}$ and
$dz$. A high value of $M_{res}$ would be required to preserve the binarity of
the code  (eq.~\ref{eq:2_3}) and to control the computational cost 
implied by a very high time resolution. However, we need $M_{res}$ to be small
enough to resolve the low mass halos presumably hosting the first stars. 
In addition, $M_{res}$ must be redshift dependent, 
decreasing with increasing redshift, to reproduce the EPS predictions (VHM). 
Indeed, hierarchical models predict less massive halos at higher redshifts. 
The halo mass which, at a given redshift $z$ corresponds to a virial equilibrium 
temperature $T_{vir}$ can be approximated as, 
\begin{equation}
M(T_{vir},z)\sim10^{8}\Msun\left(\frac{10}{1+z}\right)^{3/2}\left(\frac{T_{vir}}{10^{4}{\rm K}}\right)^{3/2}
\end{equation}
(for an exact expression see Barkana \& Loeb 2001).
Given the above redshift dependence, taking $M_{res} \propto M(T_{vir},z)$ 
appears  physically motivated. 
For reasons that will be extensively motivated in the next Section,
our analysis focuses on halos with mass $M \geq
M(T_{vir}=10^{4}$K$,z)\equiv M_{4}(z)$. Thus, we have chosen a 
resolution mass $M_{ res}(z)=M_{4}(z)/10$ which enables to 
follow the history of all MW        progenitors having $M \geq M_{4}(z)$ 
up to $z=20$.

The time-step $dz$ is empirically selected in order to obtain a good agreement 
between the EPS predictions and numerical results. We started by 
considering $820$ time-steps logarithmically spaced in expansion factor 
between $z=0$ and $z=20$. However, since the number of halos
in the high mass range was found to exceed the EPS predictions, 
we have reduced the time-step by a factor $5$ within the redshift 
interval $8<z<12$.  
 
Finally, we need to specify the assumed MW        dark matter halo mass.
Following Binney \& Merrifield (1998), we have assumed $M_{ MW}=10^{12}M_{\odot}$. 
This values likely represents a lower limit to total halo mass, being estimated
through observations of the most distant galactic satellites currently detected.

\begin{figure*}
  \centerline{\psfig{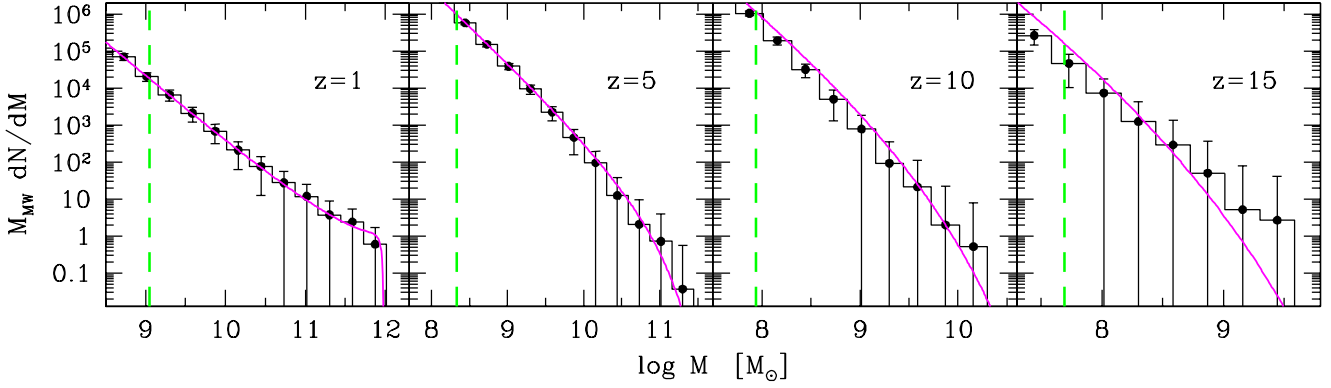}}
  \caption{Number of progenitors of a MW parent halo 
    of mass $M_{ MW}=10^{12} M_{\odot}$ as a function of mass
    at different redshifts. 
    Histograms represent averages over 200 realizations of the merger
    tree, with errorbars indicating Poissonian error on the counts in
    each mass bin. Solid lines show the predictions of EPS theory;
    dashed lines indicate the values of $M_{4}(z)$ at the
    corresponding redshift.} 
  \label{fig:1}
\end{figure*}

Comparison between the code results and EPS predictions is shown in
Fig.~\ref{fig:1}, where the mean number of Milky
Way halo progenitors is plotted at different redshifts as a function of
mass resulting from the average of $200$ different Monte Carlo realizations of the merger
tree.  At each redshift, the agreement with EPS is satisfactory down to the lowest 
halo masses $M \geq M_{4}(z)$, \ie for all the objects of interest to our study. 
Note that, as expected for $\Lambda$CDM models, the typical
population mass decreases with increasing redshift and few
collapsed halos exist at $z=15$, most of the mass being below the
resolution mass and therefore diffused in the GM.

%%%%%%%%%%%%%%%%%%%%%%%%%%%%%%%%%%%%%%%%%%%%%%%%%%%%%%%%%%%%%
%%%%%%%%%%%%%%%%%%%%%%%%%%%%%%%%%%%%%%%%%%%%%%%%%%%%%%%%%%%%%

\section{Star formation and feedback}

In hierarchical models for galaxy formation the first star-forming
halos are predicted to collapse at redshift $z\sim 20$, having masses 
$M\sim 10^6\Msun$ and virial temperatures $T_{ vir}<10^4$~K. The
neutral gas in these minihalos cannot cool via atomic hydrogen 
and relies on the presence of molecular hydrogen, H$_2$, to 
cool and collapse, ultimately forming stars. The role of
minihalos in cosmic evolution has been the subject of a number
of studies, suggesting that their star formation activity is
easily suppressed by radiative and mechanical feedback 
(see Ciardi \& Ferrara 2005 for a thorough review).
Empirical evidence for this suppression has been recently 
provided by the 3-yr WMAP data.  The downward revision of the measured 
optical depth to electron scattering of $\tau = 0.09 \pm 0.03$ (Page \etal 2006) 
is consistent with models in which ionizing sources populate halos 
down to a virial temperature of $T_{ vir} = 10^4$~K, requiring 
strong suppression of star formation in earlier minihalos 
(Haiman \& Bryan 2006). 

Consistent with these findings, we have tracked the star formation
history and chemical enrichment of the Galaxy down to progenitor
halos with masses $M_4(z)$ and we have assumed that at the initial
redshift of $z\sim 20$ the gas in these halos is still of primordial
composition. This is the result of the effect that any given
source photodissociates molecular hydrogen on scales much larger
than those affected by its metal enrichment (Scannapieco et al. 2006). 
In any star-forming halo and at each time-step, stars are assumed to
form in a single burst; the total mass of stars formed is proportional
to the available gas mass with a redshift-dependent global efficiency 
$f_{*}(z)$, $M_{*}=f_{*}M_{gas}$. The initial gas content of dark
matter halos is equal to the universal ratio,
$M_{gas}=(\Omega_{b}/\Omega_{m})M$ which decreases after each star
formation event (see eq.~\ref{eq:MgasOUT}). 

\subsection{Star formation efficiency} 
Star formation in gas clouds occurs in a free-fall time
$t_{ff}=\left({3\pi}/{32G\rho}\right)^{1/2}$ where
$G$ is the gravitational constant and $\rho$ the total (dark+baryonic) mass density
inside the halo (assumed to be 200 times denser than the background). Since, at each 
redshift, the Monte Carlo time step $\Delta t$ (corresponding to $dz$) 
is $\Delta t \ll t_{ff}(z)$, it is possible to accurately sample time variations of the 
global star formation efficiency using the following approximation: 
\be\label{eq:fstar} f_{*}(z)=\epsilon_{*}\frac{\Delta t(z)}{t_{ff}(z)},
\ee
where $\epsilon_{*}$, physically corresponding to the ``local" star formation 
efficiency, represents a free parameter of the model. 

\subsection{Pop III and Pop II/I IMFs}
We have assumed that the stellar IMF depends on the initial metallicity
of star forming clouds, in agreement with the critical metallicity scenario.
A star forming halo with an initial metallicity $Z\leq Z_{cr}$
will be referred to as a Pop III halo\footnote{We define Pop~III as
  all the stars with $Z\leq Z_{cr}$. In addition, Pop~III stars are
  assumed to be massive if $Z_{cr}>0$, and distributed according to a Larson
  IMF if $Z_{cr}=0$ (see Sec.~6.1)}, and it will host Pop III stars
with masses within the range of pair instability SNe (SN$_{\gamma\gamma}$, 
$140\Msun\leq m \leq 260\Msun$). Indeed, these violent explosions are predicted to
provide the dominant contribution to metal enrichment at the lowest 
metallicities, releasing roughly half of their progenitor mass in heavy 
elements and leaving no remnants. 
This means that all the initial progenitor mass is returned to the 
surrounding medium and that gas pollution by
these objects is very strong with respect to the contribution of
all stars with $m<40M_{\odot}$ enriching the ambient medium through mass loss or
Type II SN (SNII) explosions. We have assumed  a reference value of $m_{PopIII}=200 M_{\odot}$ for
the mass of Pop III stars, but we have also explored the implications of adopting 
the two extreme values of $140 M_{\odot}$ and $260 M_{\odot}$ in the
Appendix.

Conversely, if the initial metallicity exceeds the critical value, $Z > Z_{cr}$,
the host halo is referred to as a Pop II/I halo and the stars are assumed to form
according to a Larson IMF: 
\be
\Phi(m)=\frac{dN}{dm}\propto m^{-1+x}\exp(-m_{cut}/m),
\label{eq:Larson}
\ee
with $x=-1.35$, $m_{cut}=0.35 \Msun$ and $m$ in the range $[0.1-100] \Msun$ (Larson
1998). This expression represents a modification of the Salpeter law
which reproduces the observed present-day stellar population
for $m>1~\Msun$; in the low-mass limit, in fact, the IMF behavior is
still very uncertain because of the unknown mass-luminosity relation for
the faintest stars. The Larson IMF matches the Salpeter
law for $m>1~\Msun$ while the cutoff at $m \sim 0.35 \Msun$ can explain
the absence of brown dwarfs in the observed stellar population.

\subsection{Instantaneous Recycling Approximation}
Very massive Pop III stars are characterized by a fast evolution,
reaching the end of their main sequence phase in $3-5$ Myr.
Conversely, the broad mass range which characterizes Pop II/I stars
implies a wide range of stellar lifetimes, $\tau_{sl}$, which vary
from a few Myr to several Gyr.
In our model, we have assumed the {\it Instantaneous Recycling 
Approximation} (IRA, Tinsley 1980), according to which stars are divided in 
two classes: those which live forever, if their lifetime is longer 
than the time since their formation $\tau_{sl}>t(0)-t(z_{form})$; 
and those which die instantaneously, eventually leaving a remnant, 
if $\tau_{sl}<t(0)-t(z_{form})$. The transition mass between the 
two possible evolutions, or turn-off mass $m_1(z)$, has been 
computed at any considered redshift. All stars having mass 
$m < m_{1}(z)$ represent stellar fossils which can be observed 
today. The turn-off mass is an increasing function of
time since $[t(0)-t(z_{form})] \to 0$ when $z_{form} \to 0$; in this
limit of course $m_{1}\to 100\Msun$ \ie all the stars are still alive. 
Using the IRA approximation at each time-step, we can compute 
the number of stellar relics per unit stellar mass formed: 
\be\label{eq:Nstar}
N_{\ast}={\frac{\int_{0.1M_{\odot}}^{m_{1}(z)}{\Phi(m)dm}}{\int_{0.1M_{\odot}}^{100M_{\odot}}m\;\Phi(m) dm}},
\ee
and the equivalent mass fraction in these stars
\be\label{eq:f_mstar}
f_{m\ast}=\frac{\int_{0.1 M_\odot}^{m_1(z)}m\;\Phi(m)dm}{\int_{0.1M_\odot}^{100M_\odot}m\;\Phi(m) dm}.
\ee
By definition, there are no stellar fossils of Pop III stars in the SN$_{\gamma\gamma}$ progenitor mass range.

%%%%%%%%%%%%%%%%%%%%%%%%%%%%%%%%%%%%%%%%%%%%%%%%%%%%%%%%%%%%%%%%%%%%%%
\begin{figure*}
\centerline{\psfig{figure=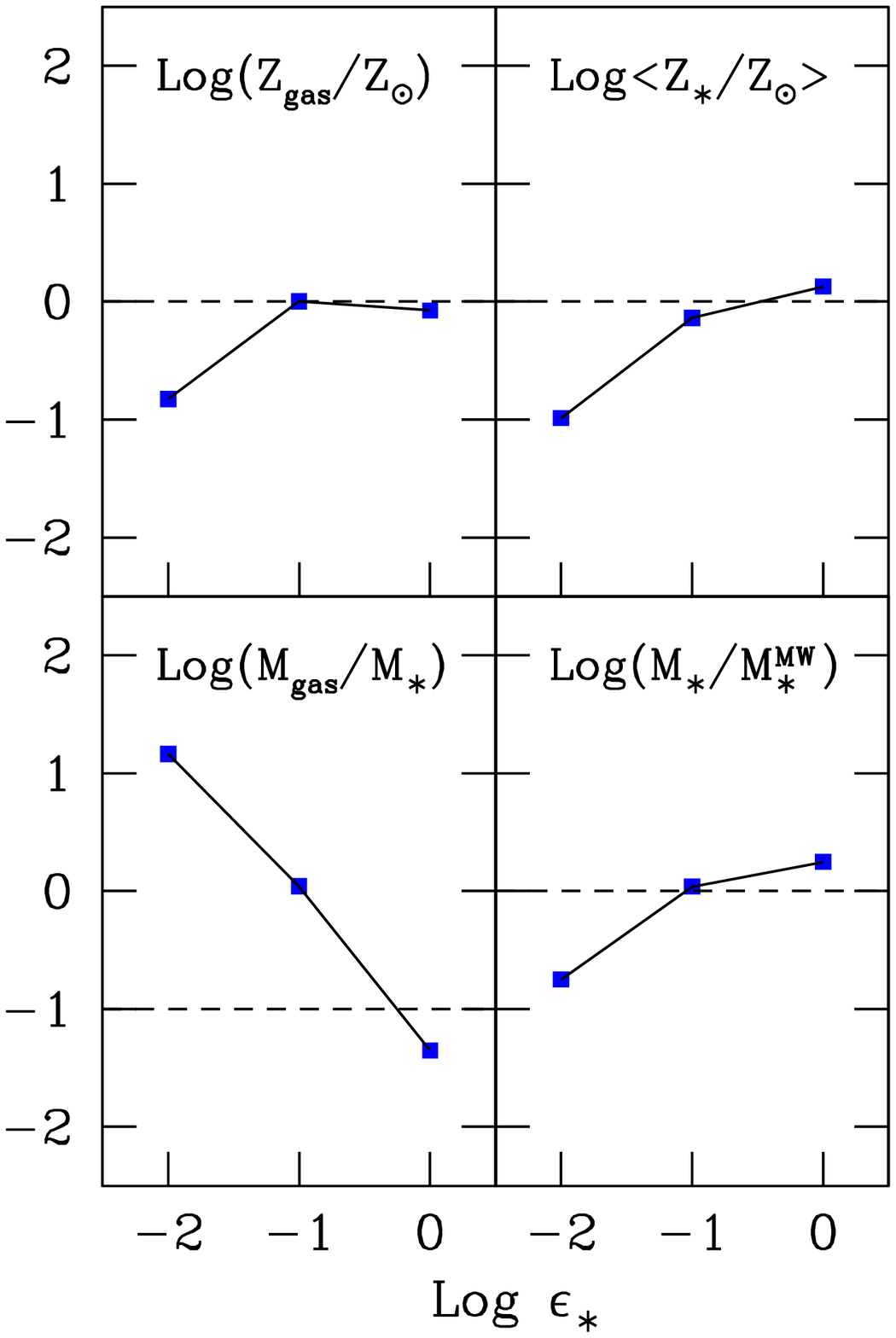,width=7.5cm,angle=0}
\psfig{figure=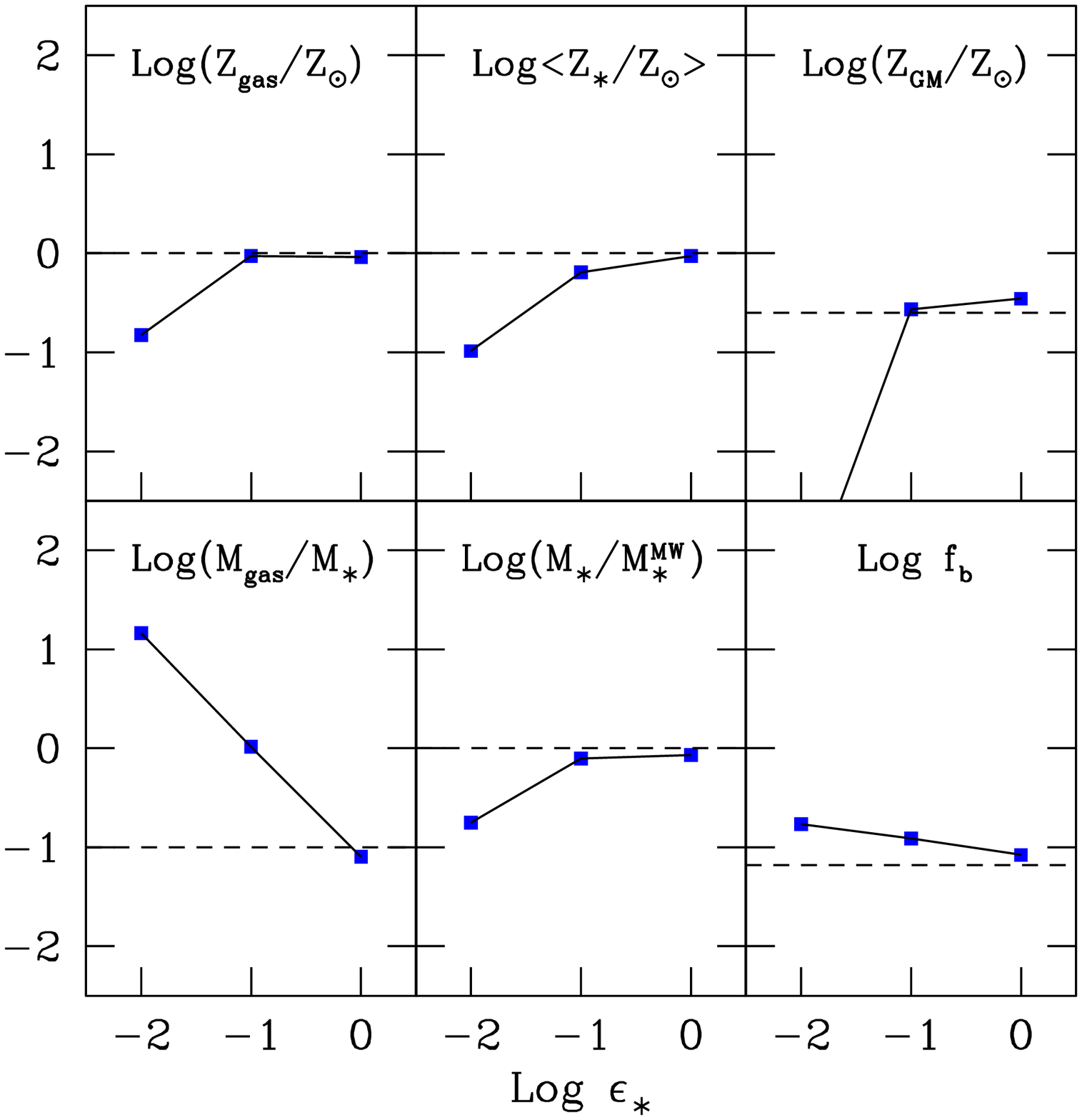,width=7.5cm,angle=0}}
\caption{Global properties of the MW. Squared points are the
  model results (200 realizations) as a function of the star formation efficiency, $\epsilon_{\star}$.
  Error bars are not shown because negligible in the selected scale. Dashed lines represent the 
  (typical) observed values. {\it Left panel}: no-feedback model
  ($\epsilon_{w}=0$); {\it Right panel}: feedback model with 
  $\epsilon_{w}=0.2$.} 
\label{fig:2}
\end{figure*} 
%%%%%%%%%%%%%%%%%%%%%%%%%%%%%%%%%%%%%%%%%%%%%%%%%%%%%%%%%%%%%%%%%%%%%%
\subsection{Nucleosynthetic products}
Massive stars can lose mass and heavy elements through stellar winds and supernova explosions. 
Using the IRA approximation, we can compute the {\it yield}, i.e. the mass 
fraction of metals produced per unit stellar mass formed,
\be\label{eq:yield}
Y=\frac{\int_{m_{1}(z)}^{100 M_{\odot}}m_{Z}(m,Z)\,\Phi(m)\,\mathrm{d}m}{\int_{0.1M_{\odot}}^{100M_{\odot}}m\;\Phi(m)\mathrm{d}m},
\ee
as well as the {\it returned fraction}, or the stellar mass fraction returned 
to the gas through winds and SN explosions:  
\be\label{eq:returned}
R=\frac{\int_{m_{1}(z)}^{100M_{\odot}}(m-w_{m}(m))\,\Phi(m)\,\mathrm{d}m}{\int_{0.1M_{\odot}}^{100M_{\odot}}\Phi(m)\,m\;\mathrm{d}m}.
\ee
The quantity $m_{Z}(m,Z)$ represents the mass of metals produced by a star with initial mass $m$ and metallicity $Z$, 
and $w(m)$ is the mass of the stellar remnant. 
Non-rotating Pop III stars in the SN$_{\gamma\gamma}$ domain return all 
their gas and metals to the surrounding medium, \ie $R=1$. We have used
for these stars the results of Heger \& Woosley (2002).
It is interesting to note that although the total metal yield is independent of the progenitor mass and equal to 
$Y=0.45$, the iron yield strongly depends on mass, being $Y_{Fe}=(2.8\times 10^{-15}, 0.022, 0.45)$ 
for $m=(140, 200, 260) M_\odot$. 
As for Pop~II/I stars  we have used the grid of models by van den Hoek \& Groenewegen (1997) for intermediate 
($0.9 M_\odot < m < 8 M_\odot$) mass stars and Woosley \& Weaver (1995) for SNII ($8 M_\odot < m < 40 M_\odot$), 
linearly interpolating among grids of different initial metallicity when necessary. 
We have also followed the evolution of individual elements relevant to the present study, Fe and O. 

\subsection{Mechanical Feedback}
SN explosions may power a wind which, if sufficiently energetic, may
overcome the gravitational pull of the host halo leading to expulsion of
gas and metals into the surrounding GM. This mechanical feedback has important
implications for the chemical evolution along the merger tree, as the nucleosynthetic
products of the first stars can be efficiently ejected out of the shallower potential
wells of the first star forming halos and then be accreted by neighboring halos,
thus being incorporated into the next stellar populations.  

To model mechanical feedback, we compare the kinetic energy injected by SN-driven winds 
\be
\label{eq:Esn}
E_{SN}=\epsilon_{w} N_{SN} \langle E_{SN}\rangle,
\ee
and the binding energy of the host halo with mass $M$,
\[
E_{b}=\frac{1}{2}\frac{GM^{2}}{r_{vir}}= 5.45\times{10^{53}}{\rm erg}\left(\frac{{M}_{8}}{h^{-1}}\right)^{5/3}\left(\frac{1+z}{10}\right){h}^{-1}
\]
\noindent
where $M_{8}={M}/{10^{8}M_{\odot}}$ (Barkana \& Loeb 1991).
In the first equation, $\epsilon_{w}$ is wind mechanical efficiency 
(i.e. the fraction of explosion energy 
converted into kinetic form); it represents the second free
parameter (the first is $\epsilon_{\star}$, see eq. \ref{eq:fstar}) of the model.
$N_{SN}$ is the number of SNe in the burst and $\langle E_{SN}\rangle$ is the 
average explosion energy, which we take to be equal to $2.7\times 10^{52}$~erg for SN$_{\gamma\gamma}$ and to 
$1.2\times 10^{51}$~erg for SNII.  
We assume that ejection takes place when $E_{SN}>E_{b}$, and the gas is retained otherwise. We compute 
the ejected fraction of gas and metals as 
\be\label{eq:Alpha}
\alpha_{ej}=(E_{SN}-E_{b})/(E_{SN}+E_{b}).
\ee 
Hence, the ejection fraction is directly proportional to the SN energy provided it is larger than
the binding energy.  Note that, according to this simple prescription, gas and metals are ejected 
with same efficiency. This might not necessarily be the case, as shown by several authors (Mac Low \& 
Ferrara 1999; Fujita et al. 2004); we neglect this complication in this work.

Due to mechanical feedback, the mass of gas and metals in a halo can decrease 
substantially. Following a star formation burst and mechanical feedback,
the mass of gas left in the halo, $M_{gas}$, which represents the reservoir
for subsequent star formation events, is related to the initial gas mass, $M_{gas}^{in}$, and the
stellar mass, $M_\star$, by 
\[
M_{gas}=[M^{in}_{gas}-M_\star+RM_{*}](1-\alpha_{ej})=
\]
\be
=M^{in}_{gas}(1-f_{*}+Rf_{*})(1-\alpha_{ej}), 
\label{eq:MgasOUT}
\ee
Similarly, the final mass of metals can be written as,
\[
M_{Z}=[M^{in}_{Z}(1-f_{*})+YM_{*}](1-\alpha_{ej})=
\]
\be
=M^{in}_{gas}(Z_{in}(1-f_{*})+Yf_{*})(1-\alpha_{ej}).
\label{eq:Zout}
\ee

\subsection{Metal mixing}
We have reconstructed the star formation and chemical enrichment
history of our Galaxy applying iteratively eqs.~(\ref{eq:MgasOUT})-
(\ref{eq:Zout}), together with eqs.~(\ref{eq:Nstar})-(\ref{eq:Alpha}), 
along the hierarchical merger tree. It is assumed that during a merger
event the metal and gas content of two distinct progenitor halos are 
perfectly mixed in the ISM of the new recipient halo. 
Similarly, metals and gas ejected into the GM are assumed to be 
instantaneously and homogeneously mixed (we refer to this
approximation as ''perfect mixing'') with the gas residing in that component.
The filling factor $Q$ of the metal bubbles inside the volume corresponding
to the size of the MW halo today, gives an estimate of the validity
of the latter assumption. The perfect mixing approximation
is verified when
\be
Q(z)={\big(\frac{R_b(z)}{\langle \lambda (z)\rangle}\big)}^3 > 1
\ee
where $R_{b}$ is the bubble radius and 
\be
\langle \lambda (z)\rangle=\frac{V_{MW}(z)}{N_h(z)}=\frac{V_{MW}(0)(1+z)^3}{N_h(z)}
\ee
is the average mean halo separation within the proper MW volume
$V_{MW}(z)$, having assumed $V_{MW}(0)\sim 1$~Mpc; $N_h(z)$ is the
total number of halos at redshift $z$, averaged over 200 realizations of the
merger tree. The value of $R_b$ can be estimated from a Sedov-Taylor
blastwave solution:      
  \be  
R_b(z)=\big[\frac{E(z)}{\langle \rho_b(z)\rangle}\big]^{1/5} t_H^{2/5}(z)
\nonumber
\ee
where $E(z)$ is the energy released by SN explosions within
each halo, $\langle \rho_b(z) \rangle$ is the mean GM density and
$t_H(z)$ is the Hubble time. If we assume that i) star forming
halos have $M_h(z) = M_4(z)$ and ii) $E(z) = E_0
f_*(z)\Omega_b/\Omega_m M_4(z)$ where $E_0 = 1.636\times
10^{49} erg/\Msun$ is the Pop~II explosion energy per unit stellar
mass formed, we find that $Q>1$ when $z<11$. Such limit implies that
regions with $Z=0$ are no longer present beyond that epoch.
Additional discussion on this issue is given in Sec.~5.2 and Sec.~7.

\section{Model calibration}

%%%%%%%%%%%%%%%%%%%%%%%%%%%%%%%%%%%%%%%%%%%%%%%%%%%%%%%%%%%%%%%%%%%%%%%%%%%%
\begin{table*}\label{table1}
  \begin{center}
    \begin{tabular}{|c|c|c|c|c|c|c|}\hline
      \hline 
       model & $Z_{gas}/Z_{\odot}$ & $<Z_{\ast}/Z_{\odot}>$ & $Z_{GM}/Z_{\odot}$ &
      $M_{gas}/M_{*}$  &  $M_{\ast}/M^{MW}_{\ast}$  &  $f_{b}$ \\
      \hline 
      no feedback: $\epsilon_{*}=0.5$ , $\epsilon_{w}=0$ & $1.261$ & $1.13$ & $--$ & $0.115$ 
      & $1.71$ & $--$ \\
      \hline
      feedback:    $\epsilon_{*}=0.7$ , $\epsilon_{w}=0.2$ &
       $0.93$ & $1.01$ & $0.41$ &
       $0.11$ & $0.85$ & $0.085$ \\
      \hline
       observed values & 1 & 1 & 0.25 & 0.1 & 1 & 0.066 \\
      \hline
    \end{tabular}
  \end{center}
  \caption{Averaged values of the global properties of the MW derived using
  the two best-fit models: no-feedback $\epsilon_{*}=0.5$,
  $\epsilon_{w}=0$, and feedback $\epsilon_{*}=0.7$,
  $\epsilon_{w}=0.2$. The values have been obtained averaging over
  200 realizations of the merger tree. For comparison, we also quote
  the indicative observed values (see text).} 
\end{table*}
%%%%%%%%%%%%%%%%%%%%%%%%%%%%%%%%%%%%%%%%%%%%%%%%%%%%%%%%%%%%%%%%%%%%%%%%%%%%%%%

Although we have kept our model as simple as possible, it includes a 
number of relatively poorly known (albeit important) physical processes that
need to be empirically calibrated. To this aim we have used the observed properties 
of the Milky Way as a benchmark to fix the best values of the two model free parameters,
$\epsilon_{*}$ and $\epsilon_{w}$. In particular, we have compared the
results of the simulations at redshift $z=0$ with the following observations:

\begin{itemize}
\item Gas metallicity. The Galactic disk has a mean metallicity $Z_{gas} \simeq Z_{\odot}$. 
  As this component contains the majority of the MW        gas, we use this
  value as representative of the whole system. 
\item Stellar metallicity. The typical mean value quoted for this quantity is $Z_{*} \simeq \Zsun$. 
  This value has been derived by weighting the mean metallicity of each galactic
  stellar component with its corresponding mass. 
\item Stellar mass.  Contributions to the stellar mass come from the  
  disk ($M^{disk}_{*}\approx (4-6) \times 10^{10}\Msun$), the bulge  
  ($M^{bulge}_{*}\approx (0.4-1)~10^{10}\Msun$) and the halo  ($M^{halo}_{*}\approx (0.2-1)~10^{10}\Msun$) 
components (Dehennen \& Binney, 1998; Brown et al. 2005), summing up in a total mass of
$M_{\star}\approx 6\times 10^{10}\Msun$. 
\item Gas-to-stellar mass ratio, $M_{gas}/M_{*}=0.13$.
  The mass of gas has been derived using the observed mass of HI
  and HII regions of the Galaxy, $M_{gas} = M_{HI}+M_{HII}\sim
  (6 + 2)\times 10^{9}\Msun\sim 8 \times 10^{9}\Msun$
  (Stahler \& Palla 2004). 
\end{itemize}
Moreover, if {\it mechanical feedback} is  considered, two additional 
quantities become meaningful: 
\begin{itemize}
\item Baryon to dark matter ratio, $f_{b} =(M_\star+M_{gas})/M_{ MW} = 0.07$.  
\item GM metallicity. Although the exact value for this quantity is somewhat 
uncertain, high-velocity clouds (supposedly constituted by leftover gas from the
Galactic collapse, and currently accreting onto the disk) may be used as a reasonable template.    
For these objects, Ganguly \etal (2005) derive  [O/H]$\sim -0.66 \sim 0.25 \Zsun$, 
a value that we take as an educated guess for $Z_{GM}$. 
\end{itemize}
Obviously, in study cases in which mechanical feedback is turned off for comparison sakes, it
follows that $Z_{GM}=0$ and $f_{b}=\Omega_{b}/\Omega_{m}=0.16$. 
In principle, an acceptable fit to the above MW properties can be achieved both 
for models including mechanical feedback and neglecting it (see next Sections). 
However, the best values of the free parameters are different in the two cases. 
We will show later how it is possible to assess the role of feedback by using the 
MDF properties. Note that the calibration procedure is completely
independent of $Z_{cr}$ and Pop~III IMF.

\subsection{No Feedback Model}
In Fig. \ref{fig:2} we compare the observed global properties of the
Galaxy with the results of our model as a function of the star formation 
efficiency, $\epsilon_{*}$. We start by analyzing the no-feedback case, i.e. 
when $\epsilon_{w}=0$. A good overall agreement with the four observed
quantities is obtained for relatively high values of $\epsilon_{*} > 0.1$. 
In fact, the derived values of $Z_{gas}/\Zsun$, $Z_{*}/\Zsun$, $M_{*}/M^{MW}_{*}$ 
rapidly increase when the star formation efficiency increases from $\epsilon_{*}=0.01$ to
$0.1$, and remain nearly constant thereafter
(note that the value of $M_{gas}/M_{*}$ actually {\it decreases} with 
$\epsilon_{*}$ in this range). These trends can be understood very simply.  
The gas mass decreases due to star formation (eq.~\ref{eq:MgasOUT}), forcing the 
gas fraction to decrease with increasing $\epsilon_*$.
Thus, $M_{*}$ saturates beyond $\epsilon_{*}>0.1$ because of the
limited gas mass available. Given the proportionality 
between $M_{Z}$ and $M_{*}$ (eq.~\ref{eq:Zout}), $\langle Z_{*} \rangle$ 
shows the same behavior.  Conversely, the mass of metals in the gas is 
determined by the balance between metals synthesized and ejected by stars, 
and metals locked into newly formed stars (eq.~\ref{eq:Zout}); when 
the second effect dominates, it produce a slight decrease of $Z_{gas}$. 
The best-fit model requires a high star formation efficiency, $\epsilon_*=0.5$\footnote{
Because of the assumed star formation law (eq.~\ref{eq:fstar}),
$\epsilon_{*}=0.5$ is only the instantaneous star formation efficiency and 
does not imply that half of the gas mass is converted 
into star in a single burst. Rather, for this model $f_{*}(z)$ 
varies in the range $(0.06-4)\times 10^{-2}$ in the considered redshift range.}. 
The derived values for all the observed properties of the MW are
listed in Table~1. The values shown represent an
average over $200$ realizations of the merger tree. 
 
\subsection{Feedback Model}
In the right panel of Fig.~\ref{fig:2}, we show the global 
properties of the MW derived using a feedback model with 
$\epsilon_{w}=0.2$. For low values of $\epsilon_\star$,
the effect of mechanical feedback is negligible; similarly to the no
feedback model $f_{b}\approx \Omega_{b}/\Omega_{m} =0.16$,  and $Z_{GM}\approx 0$. 
These results are coherent with our feedback 
prescriptions (eqs.~\ref{eq:Esn} and~\ref{eq:Alpha}) since 
the efficiency of mechanical feedback depends on the number
of SN explosions and therefore on the level of star formation activity. 
As expected, $f_b$ decreases with $\epsilon_{*}$ while $Z_{GM}$ increases. 
Note that in the range  $0.1 < \epsilon_{*} < 1$  the feedback model 
closely resembles the no-feedback one, the major difference being 
the lower saturation values for the stellar mass and metallicity. 
Instead, the metallicity of the gas is nearly unaffected because 
metals and gas are ejected into the GM with the same efficiency 
(see eqs.~\ref{eq:MgasOUT} and~\ref{eq:Zout}).  The best-fit model requires 
$\epsilon_{*}=0.7$ (we stress again that this is
{\it not} the global star formation efficiency, expressed by $f_\star(z)$, see 
eq. \ref{eq:fstar}), and reproduces the MW global properties better that  
the no-feedback model (see Table~1).

\section{Model results}

\subsection{Metallicity Distribution Function}

After having ensured that our models reproduce the global properties of the MW, 
regardless of the value of $Z_{cr}$ and of the PopIII IMF,
we are now ready to explore the predictions of the calibrated models.
The result shown below are the result of averages over 200 realizations 
of the MW merger tree; in some cases we use a single representative
realization for illustration purposes.

%%%%%%%%%%%%%%%%%%%%%%%%%%%%%%%%%%%%%%%%%%%%%%%%%%%%%%%%%%%%%%%%%
\begin{figure*}
  \centerline{\psfig{figure=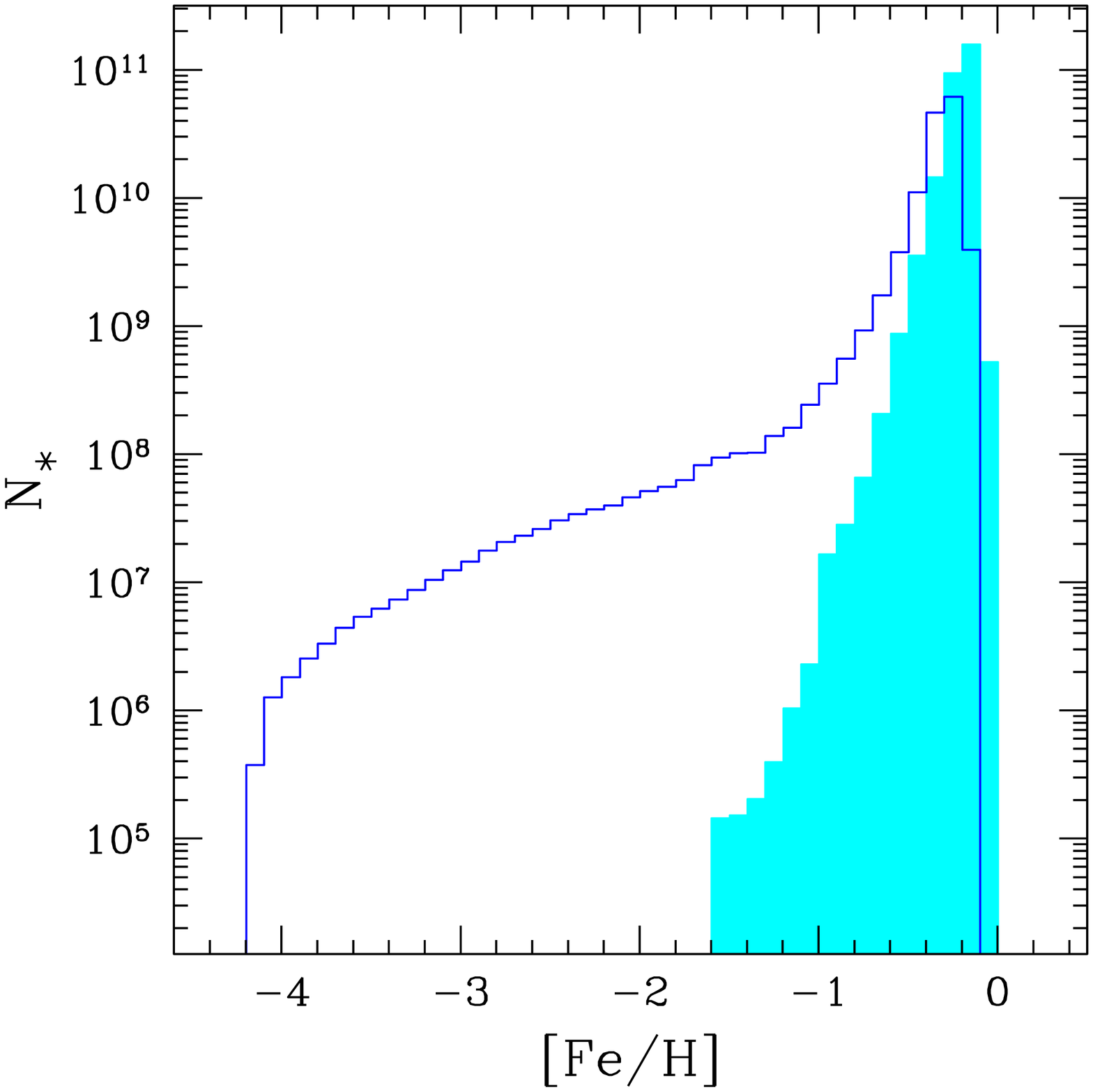,width=8.0cm,angle=0}
  \psfig{figure=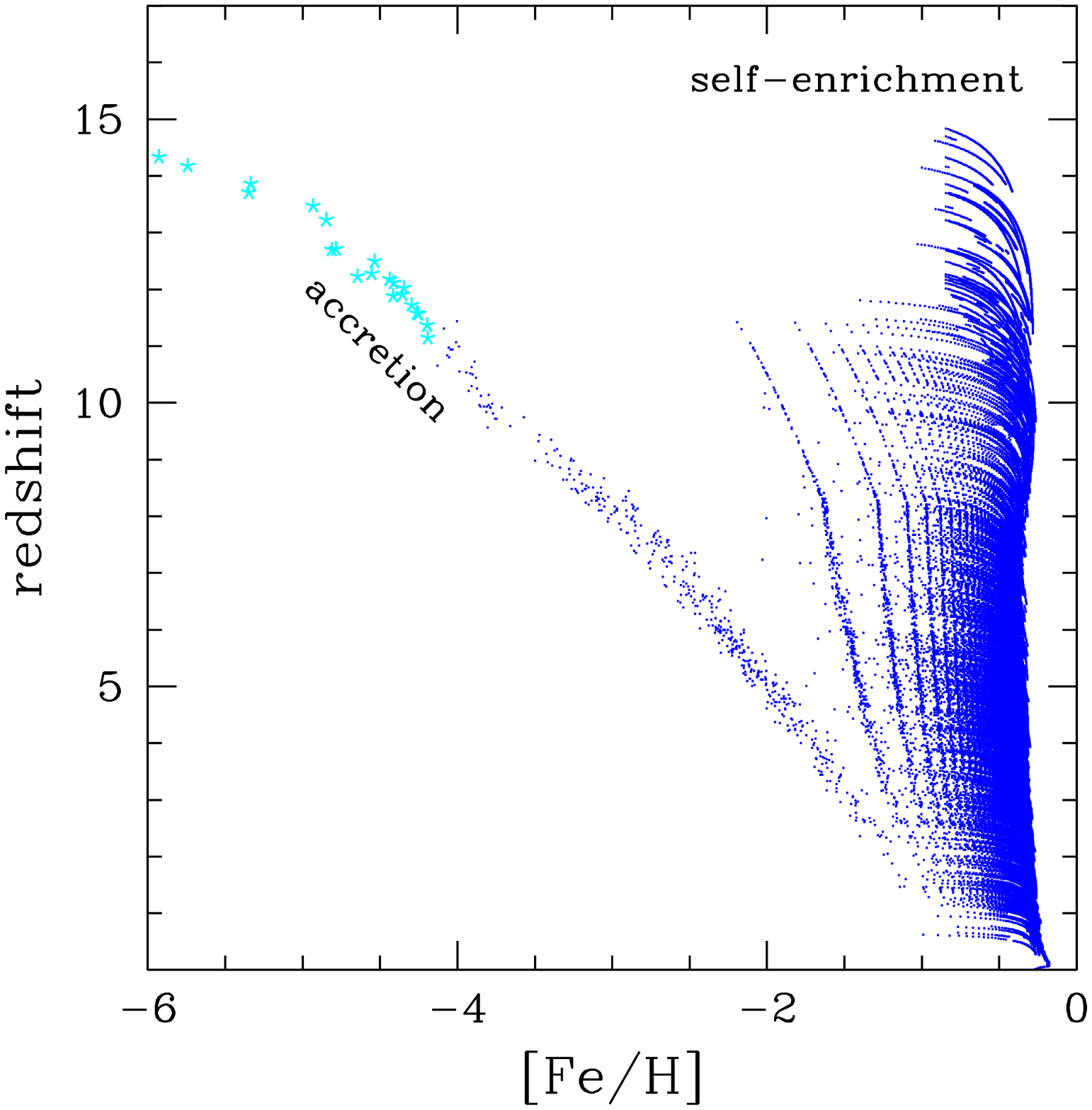,width=8.0cm,angle=0}}
  \caption{{\it Left panel:} MDFs derived using the no-feedback model 
    (shaded histogram), and the fiducial feedback model (unshaded) both summarized in Table 1. 
    Histograms have been obtained as average over 200 realizations. 
    Both models assume $Z_{cr}=10^{-4}\Zsun$ and $m_{ PopIII}=200\Msun$. {\it Right panel:} 
    Age-metallicity relation for the halos derived using the fiducial feedback
    model. The starred (solid) points represent Pop~III (Pop~II/I)
    star-forming halos. For clarity results are shown for
    a single realization.}
\label{fig:3}
\end{figure*}
%%%%%%%%%%%%%%%%%%%%%%%%%%%%%%%%%%%%%%%%%%%%%%%%%%%%%%%%%%%%%%%%%

In Fig.~\ref{fig:3} we compare the MDF (\ie the number of MW relic stars as a
function of their iron abundance, [Fe/H], a quantity commonly used as
a metallicity tracer) derived using the two best-fit
models, with and without mechanical feedback (see Table~1),
for the specific value $Z_{cr}=10^{-4}\Zsun$ and $m_{PopIII}=200\Msun$.

The first striking difference is that no stars with [Fe/H]$\lsim-1.5$ are produced 
in the no-feedback model. The reason is that the gas metallicity is
not efficiently diluted in closed box models. A simple estimate elucidates this point. 
Following a burst of Pop~III stars, the gas metallicity of the host halo is raised up to 
\be
Z=\frac{M_{Z}}{M_{gas}}=\frac{Y M_{*}}{M_{gas}}=Y f_{*}(z)\gsim 0.146\Zsun,
\label{eq:Self_en}
\ee
where we have used $Y=0.45$ and the smallest $f_{*}(z)\sim 6.5\times
10^{-3}$ value corresponding to $\epsilon_{*}=0.5$ (appropriate for
the no-feedback case).   
Note that the final metallicity depends only on the metal yield and on the star
formation efficiency. From the previous equation it is easy to see that the resulting abundance is\footnote{This relation follows from 
the definition of [Fe/H] that leads to [Fe/H]$=\log\left({{Z}}/{\Zsun}\right)+1.044+
\log\left(Y_{ Fe}/{Y_Z}\right)$, where for Pop III stars with $m=200\Msun$ the iron yield is 
approximately $Y_{Fe} = 0.05Y$.} 
\[{\rm [Fe/H]} \simeq \log\left(Z/\Zsun\right)-0.257 \gsim -1.09\]
This value is larger than the low [Fe/H] cutoff in the distribution
shown in Fig.~\ref{fig:3}. What is the reason for such discrepancy?  
In the absence of mechanical feedback, metals can only be diluted through 
mass accretion of metal-free gas and through mergers with unpolluted 
progenitors (which at most can halve the gas metallicity $Z$ if they have a 
comparable mass). However, dilution purely provided by accretion of unpolluted gas is not
sufficient to account for the low-$Z$ tail of the MDF, which extends to values 
[Fe/H] $\approx -4$, with two outliers at [Fe/H] $<-5$. 
The same conclusion is achieved using different mass values for
SN$_{\gamma\gamma}$ or assuming a Larson IMF for metal free stars though,
in both cases, the different value of the yields cause a shift in the
MDF cutoff. For example for the Larson IMF ($Z_{cr}=0$
model) $Y\sim 0.01$ and $Y_{Fe}\sim 5\times 10^{-4}$
and the cutoff moves to [Fe/H]$\sim -2.44 \gg -4$. 
A similar result is obtained for $m_{PopIII}=140\Msun$ shown in the
Appendix.\\  
In conclusion mechanical feedback is required to effectively dilute metals
inside halos independently of the $Z_{cr}$ value and of the assumed
PopIII IMF.

Heavy elements ejected by SN winds mix with the external GM and 
can be accreted by other halos, raising their metallicity. Contrary
to self-enrichment by previous starburst within the same host halo
and to hierarchical enrichment by contaminated progenitors, this
accretion-driven enrichment mode is present only when mechanical
feedback operates. Accretion of metal-enriched GM onto a halo of
primordial composition sets the initial conditions for the environment
in which most metal-poor stars will later form. Indeed, the unshaded
histogram in Fig.~\ref{fig:3} shows that when mechanical feedback is active
the MDF extends down to [Fe/H]$ \gsim \log(Z_{cr}/\Zsun)$ as low-mass 
stellar relics can be produced only in gas clouds with $Z > Z_{cr}$.

The right panel of Fig.~\ref{fig:3} shows how different enrichment modes
operate along the merger tree in a single representative realization. 
Each point denotes the metallicity of a halo at a given redshift; in
this plane halos move from left to right \ie they increase their
metallicity. The evolution of each halos is followed along its track
until it suffers a merger event. Two families of curves can be
identified, corresponding to self-enrichment and accretion mode,
respectively, as marked by the labels in the Figure.
In general, at the onset of the first burst of star formation a halo
could be either metal-free ([Fe/H] $ = -\infty$) or it could have a
non-zero metallicity if it has accreted gas from the GM previously
polluted by metals ejected from other halos\footnote{The same
conclusion applies to halos that are just about to gravitationally
 collapse and virialize.}.
In the first case, the halo will host Pop~III stars; if metals produced
are retained after their explosion the halo becomes self-enriched
and moves directly on the corresponding self-enriched track.
In the second case, instead, depending on the relative ratio of the GM
metallicity $Z$ and $Z_{cr}=10^{-4}\Zsun$ the stars form could be
either Pop~III or Pop~II. In this particular realization at $z\sim 11$ the GM metallicity 
crosses the $Z_{cr}$ value and sets the termination of the Pop~III formation epoch in
accretion-enriched halos. Such halos lie on the curve marked as
``accretion''. However the second star-formation episode will push
again these halos onto the corresponding self-enriched tracks. The
accretion-mode is crucial in order to form long-lived stars with
[Fe/H]$<-2.5$. The accretion-enriched branch ends at $z\sim 1$, when
merger events dominate the evolution of MW progenitors. The last point
of the branch corresponds to [Fe/H] $\sim -1$, reflecting the
metallicity of the GM at the same redshift (see Fig.~\ref{fig:4}).   

The analysis of this Figure helps interpreting the MDF. Our model  
predicts that stars with [Fe/H]$<-2.5$ form {\it only} in halos enriched through 
the accretion-mode. Conversely, iron-rich stars, with [Fe/H]$>-1$, 
form in self-enriched halos. In the intermediate range, 
$-2.5 \le $[Fe/H]$ \le -1$, stellar relics originate from both
self-enriched and accretion-enriched halos, and their contribution
to the MDF cannot be separated. However, at a given [Fe/H],
stellar relics along the accretion-enriched branch are
systematically younger.     

Note that another possibility to increase metal dilution
along the merger tree is to relax the hypothesis that 
all Pop~III stars end up as SN$_{\gamma \gamma}$ 
($f_{\gamma \gamma}=1$). Indeed if a fraction of
Pop III stars forms with $m>260\Msun$ or $m<140\Msun$ 
($f_{\gamma \gamma}<1$) and ends up as black holes 
(Heger \& Woosley 2002), metal enrichment can be 
substantially reduced. However, Schneider \etal (2006) 
have recently studied the dependence of chemical feedback
from the parameter $f_{\gamma \gamma}$ (thus from the assumed
Pop III IMF), showing that $f_{\gamma \gamma}=1$ is currently
favored by observations of the cosmic star formation history,
high redshift number counts and cosmic reionization.

In conclusion mechanical feedback is necessary to reproduce both the
global properties of the MW and the observed range 
of [Fe/H] in halo stars. Hereafter we will refer to the
best-fit feedback model of Table~1 as our fiducial 
model.
    
\subsection{Metallicity evolution of the Galactic Medium}

\begin{figure}
  \centerline{\psfig{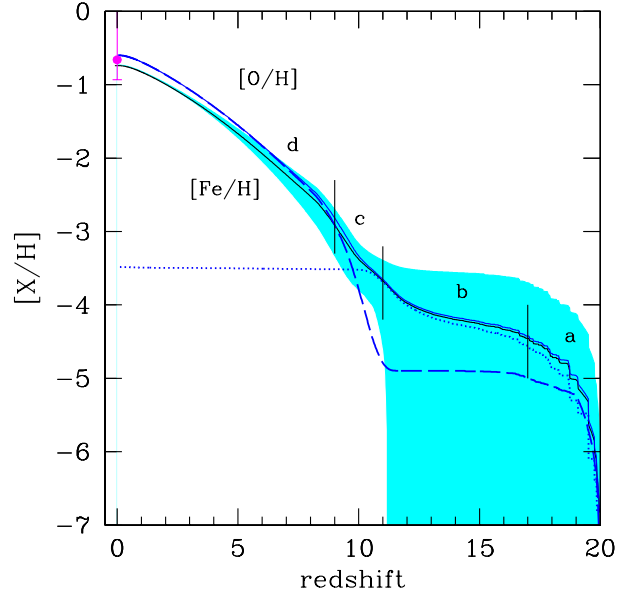}}
\caption{GM elemental abundance as a function of redshift for the fiducial feedback model
(see Table 1) and $Z_{cr}=10^{-4}\Zsun$, $m_{PopIII}=200\Msun$. Solid lines
  show the evolution of oxygen and iron abundance, averaged over
    200 realizations of the merger tree; the shaded area delimits the
    $\pm 1\sigma$ dispersion region. The [O/H] contribution by Pop~III
  (dotted line) and Pop~II stars (dashed line) are also shown
  separately. The point is the measured [O/H] in high-velocity clouds
  (Ganguly et al. 2005).} 
\label{fig:4}
\end{figure}
	
In Fig.~\ref{fig:4} we show the evolution of GM iron and oxygen abundances 
together with the specific contribution to [O/H] by Pop~III and Pop~II stars.
For reference we also plot the observed [O/H] value measured in high-velocity clouds (Ganguly
et al. 2005) which are taken as an indicator of leftover GM in the
MW halo. These results refer to our fiducial model with
$Z_{cr}=10^{-4}\Zsun$ and $m_{PopIII}=200\Msun$. 
The predicted [O/H] at $z=0$ is in perfect agreement with the data.
The scatter of the [Fe/H] distribution increases with $z$ and becomes
very large at $z\sim 11$ as a result of the large fluctuations in the
merging history (see Fig.~1) and enrichment of halos among different
realizations at $z>10$. This result is consistent with the fact that
$Q=1$ at $z=11$ (see Sec.~3.6) and therefore no unpolluted regions are
expected below that redshift. The scattering induced by the
stochastic nature of the merger tree process is similar, although
different in nature, with respect to the abundance fluctuations
produced by inhomogeneous mixing, a process not describe by our
perfect mixing approximation. This similarity can be appreciated by
comparing our results with those found by Mori \& Umemura (2006) using a hybrid
N-body/hydrodynamic code. These authors conclude that the metallicity is
highly inhomogeneous during the early phases of galaxy formation ($t\leq
0.1$~Gyr for a $10^{11}\Msun$ galaxy) and becomes more uniform at
lower redshifts. 

We can see that Pop~III stars dominate the GM enrichment
for $9<z<20$ while Pop~II stars are the dominant enrichment channel for $0<z<9$.
The GM enrichment history is the result of 4 phases in which
different physical mechanism dominate; these phases are marked in
Figure~\ref{fig:4}. The first one (phase a), in the redshift range $17\lsim
z\lsim 20$, is characterized by a very rapid increase of the
elemental abundances. 
Given the small binding energy of the high-redshift, low-mass halos,
both SN$_{\gamma\gamma}$ and the less energetic SNII explosions can 
overcome the gravitational pull of parent halos, contributing to the 
GM enrichment.
As the typical halo mass scale grows, as a result of hierarchical
formation process, metals produced by SNII are retained inside
galaxies and only SN$_{\gamma\gamma}$ regulate the GM enrichment
(phase b). However the abundance grow is limited by the fact that an
increasing fraction of the GM is polluted above the $Z_{cr}$, hence
quenching the formation of Pop~III stars which eventually comes to an
end at the beginning of phase (c) when the whole GM is enriched to
$Z>Z_{cr}$. Thanks to the higher star formation efficiency of larger
halos, Pop~II stars, during this phase, can eject their heavy elements.
Their ability to do so, though, decreases with time because of the
counteracting effect of the gravitational potential of larger galaxies
that are forming below $z=5$ (phase d).
Note that the GM bears the nucleosynthetic signatures of (predominantly) 
Pop~III stars at $z>10$; their abundance pattern becomes more and more elusive at
later times when Pop~II stars start to dominate the enrichment.
This physical picture remains qualitatively unaffected by changing the
value of $Z_{cr}$ although the relative duration and amplitude of the
different enrichment phases may vary. 
This enrichment history is particular important to interpret the
observed MDF features as we will discuss in Sec.~6 as the various
phases can be traced by the number of stars that are formed during
each of them. 

\subsection{Mass ejection from progenitor halos}

\begin{figure}
  \centerline{\psfig{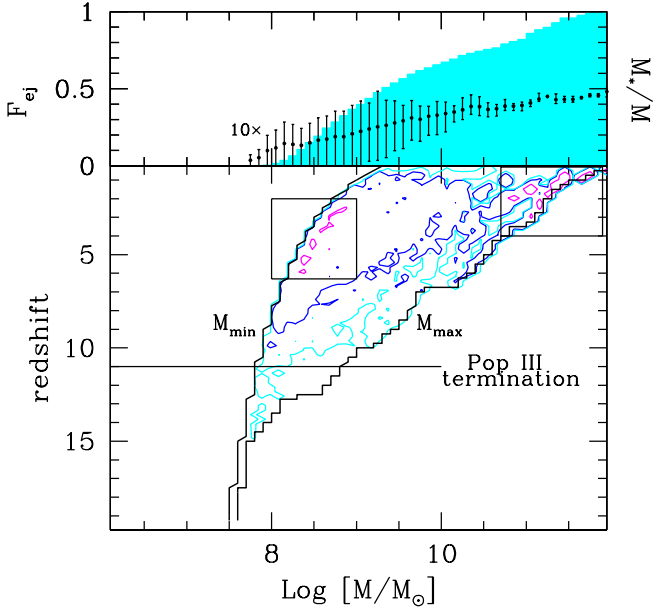}}
  \caption{{\it Lower panel}: Ratio of metals ejected by halos
    $M_{ej}$ as a function of their mass and redshift with respect to the total
    amount of metals predicted in the GM at $z=0$,
    $M^{tot}_{ej}$. Curves represent
    $M_{ej}/M^{tot}_{ej}=5\times 10^{-(4,3,2)}$ isocontours; also shown are
    $M_{min}$ and $M_{max}$. The Pop~III stars termination epoch ($z=11$) is shown
by the horizontal line. The two rectangles identify the position of
    the maxima (see text). {\it Top panel}: Cumulative fractional
    contribution, $F_{ej}$, to $M^{tot}_{ej}$ integrated over redshift by halos
    with different mass (shaded area); points with associated $\pm
    1\sigma$ error bars represent stellar-to-total mass ratios in the corresponding mass bin
    (multiplied by 10).}
\label{fig:5}
\end{figure}

%%%%%%%%%%%%%%%%%%%%%%%%%%%%%%%%%%%%%%%%%%%%%%%%%%%%%%%%%%%%%%%%%%%%%
\begin{figure*}
  \centerline{\psfig{figure=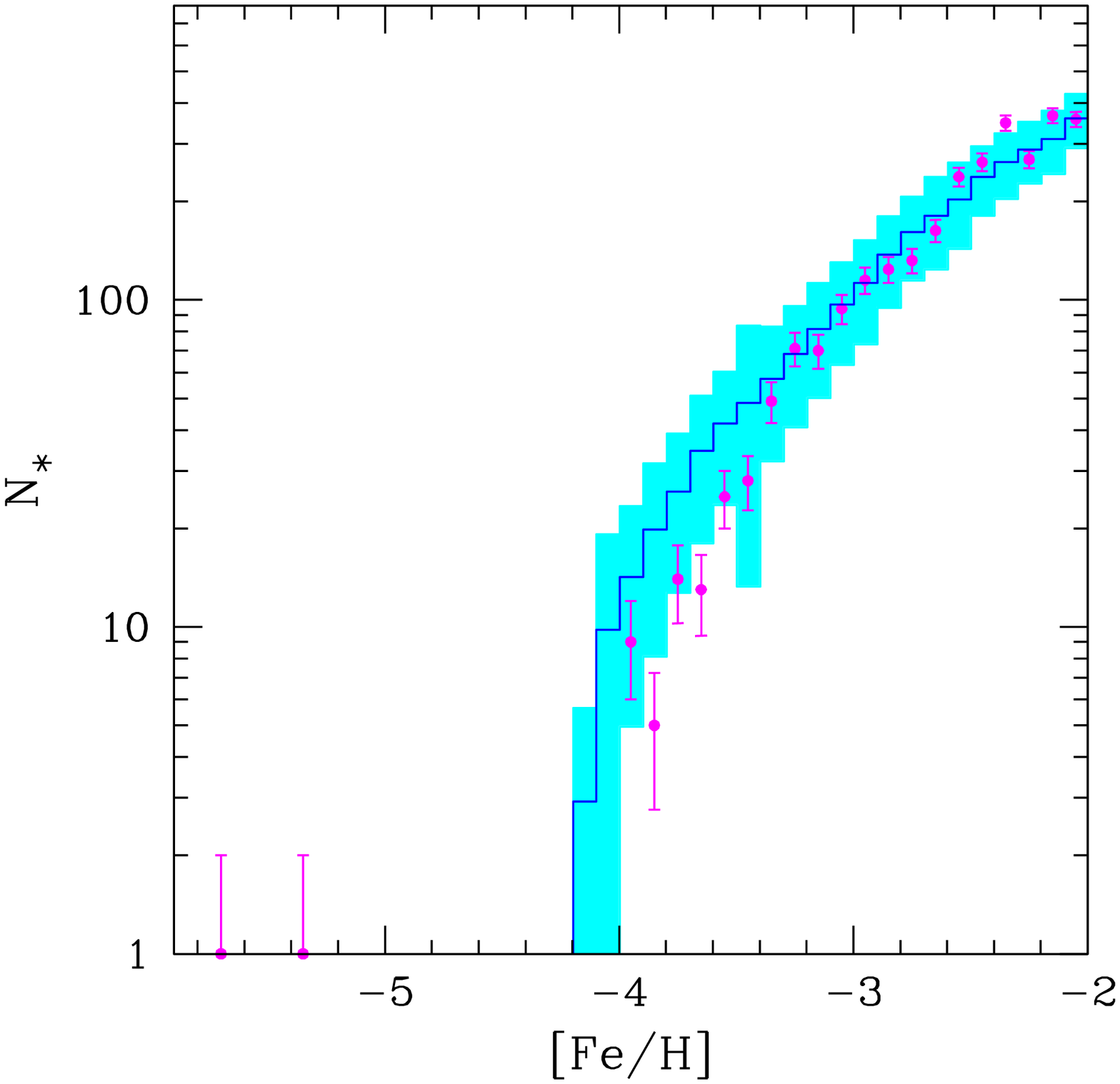,width=8cm,angle=0}
    \psfig{figure=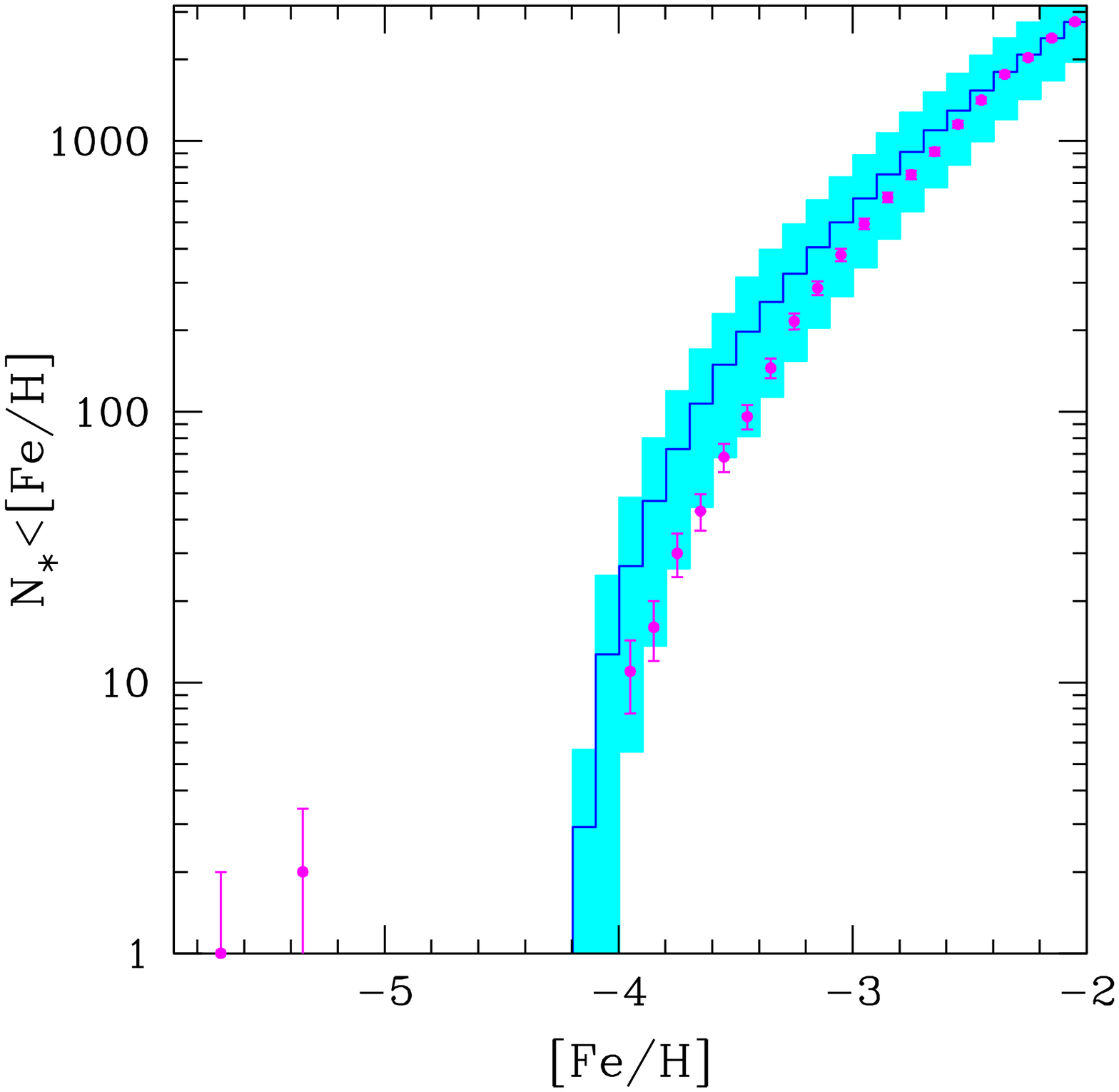,width=8cm,angle=0}}
  \caption{{\it Left panel}: Comparison with the MDF
    for the galactic halo stars ob(served by Beers \& Christlieb (2006)
    with the inclusion of the two hyper-metal poor stars (Christlieb
    et al. 2002; Frebel et al. 2005) (points) and the MDF obtained for
    the fiducial feedback model with $Z_{cr}=10^{-4}\Zsun$,$m_{ PopIII}=200M_{\odot}$ 
    (histogram). The histogram is the average value of the MDF
    over 200 realizations of the merger tree and re-normalized to the
    number of observed stars with [Fe/H]$\leq -2$. The shaded area represents $\pm
    1\sigma$ Poissonian errors. {\it Right}: The same results plotted in terms of the 
    {\it cumulative} number of stars below a given [Fe/H].}
\label{fig:6}
\end{figure*} 
%%%%%%%%%%%%%%%%%%%%%%%%%%%%%%%%%%%%%%%%%%%%%%%%%%%%%%%%%%%%%%%%%%%%

In Fig.~\ref{fig:5} we show, for a single representative realization, the
ratio of metals (or, equivalently, gas\footnote{These
  quantities are equivalent because of the assumption that the wind
  has the same metallicity as the ISM of the parent galaxy (see
  Sec.~3.5).}) ejected by halos, $M_{ej}$, as a function of their mass and
redshift with respect to the total amount of metals predicted in the
GM at $z=0$, $M^{tot}_{ej}$. In the Figure, the curve $M_{min}$ 
($M_{max}$) denotes the minimum (maximum) halo mass in which star formation can
develop, \ie $M_{min}=M_{4}(z)$. Coherently with $\Lambda$CDM
models, at high redshifts ($15<z<20$) halos have typical masses close to
$M_{min}$. As the redshift decreases,
more massive halos are produced (predominantly) via merging events which represent
the dominant formation channel for $z<6$, making $M_{max} \gg M_{min}$.
As a general rule, the GM metal enrichment is dominated for $z>2$ by
low-mass galaxies; it is only at later epochs that the contribution of larger galaxies
becomes important. In fact the distribution seen in Fig.~\ref{fig:5} is bimodal:
the first, more extended peak is found for $3<z<7$ corresponding to
halos in the mass range $M=10^{8-9}\Msun$; the second peak 
corresponds to larger galaxies $M=10^{11}\Msun$ at $z<2$.
The bi-modality can be understood as the result of the larger
$\alpha_{ej}$ of low-mass halos (due to their lower gravitational
potential) and the larger gas content of more massive ones. The first
effect dominates at high redshift; below $z=2$ the ejected mass
from larger halos becomes substantial despite their relatively low values of $\alpha_{ej}$. 
Halos in the intermediate mass range are born from a
relatively small number of merging events of gas-poor progenitors and
therefore little gas is left for further ejection. The largest halos
instead are characterized by higher baryonic fractions as they
incorporate a larger number of progenitors with different histories.
For example, halos with mass $10^{8.5}\Msun$ at $z=5$ have contributed
alone to $0.5\%$ of the metals present in the GM today. An equivalent
contribution is provided by the most massive MW progenitors in the second peak.
In the first peak a large plateau is presents around the maximum
indicating that low-mass galaxies $M<6\times 10^{9}\Msun$ play a
fundamental role for the enrichment of the GM. 

This is even more evident from an inspection of the top panel of the
same Figure where we show the cumulative contribution to
$M^{tot}_{ej}$ integrated over redshift ($F_{ej}$). The curve grows rapidly for 
$10^{8}\Msun <M< 6\times 10^{9}\Msun$ and $M>10^{11}\Msun$, with a
flatter behavior in between, reflecting the afore-mentioned bimodal
distribution. About $60\%$ of the ejected mass at $z=0$ comes from halos with
$M\lsim 6\times 10^{9}\Msun$, compared to $20\%$ contribution from the most
massive galaxies. We conclude that GM enrichment is dominated by low-mass halos 
$M\lsim 6\times 10^{9}\Msun$. This result is insensitive to the value
of $Z_{cr}$, although the relative contribution of Pop~III and Pop~II stars may vary.

As in our code we also store the information about the stellar mass
corresponding to each DM halo we can rephrase the previous results in
term of such quantity (points in the top panel of Fig.~\ref{fig:5}).
Proto-galaxies that mostly contribute to GM pollution have a typical 
stellar-to-total mass ratio $M_{*}/M\lsim 0.03$, or 
$M_{*}\lsim 2\times 10^8 \Msun$.

Note however that the scatter of the $M_{*}/M$ ratio can be
relatively large as shown by the error bars in the plot:
the dispersion is caused by the different star formation histories.
The scatter is maximum around $M=10^9\Msun$; this population is
the most numerous one through most of the MW history and they may be
produced by widely different combinations of formation processes and
virialization epochs.

From Fig.~\ref{fig:4} we have concluded that Pop~III dominate the enrichment
down to $z = 11$; from Fig.~\ref{fig:5} we can also set an upper limit to the
fraction of today's GM metals provided by Pop~III stars. At $z=11$ the
largest halo from which metals can escape is $M\leq 2\times
10^{8}\Msun$, corresponding to $F^{PopIII}_{ej}<0.04$ \ie Pop~III
stars contribute negligibly to the heavy elements currently detectable
in the Galactic environment.

\section{Data comparison}

We are now ready to compare the results of our model with the best currently available data on halo 
metal-poor stars in terms of their MDF.  Such distribution gives the number of MW relic stars 
as a function of their iron abundance, [Fe/H], a quantity commonly used as a metallicity tracer. 
In Fig.~\ref{fig:6} we compare the joint HK/HES MDF by Beers \& Christlieb (2006) 
with the simulated MDF (plotted both in differential and cumulative form) obtained using our fiducial 
model with $Z_{cr}=10^{-4}\Zsun$ and $m_{PopIII}=200\Msun$.
For completeness, we have added to the above sample the two known hyper-metal poor stars
(Christlieb \etal~2002; Frebel \etal~2005) using the new
determination by Christlieb, Bessell \& Eriksson (2006) for the iron-abundance of the star
HE0107-5240.  Due to contamination by disk stars, we have cut the data sample 
for [Fe/H]$> -2$. To compare with data, we have normalized the simulated MDF to the 
total number of observed stars with [Fe/H]$\leq -2$. Note that comparing these quantities we are
implicitly making the assumption that all simulated metal-poor stars reside
in the halo: this hypothesis is supported by the most recent N-body simulations (Scannapieco et al. 2006).

The model reproduces the observed MDF quite well, particularly for [Fe/H]$> -3.2$. A 
marginally significant deviation from the data (always within 1-$\sigma$) is seen 
at lower [Fe/H] \ie a range populated by stars the formed in accretion-enriched halos.
The two hyper-metal poor stars are not reproduced by our model. By definition in fact, 
the formation of Pop~II stars is allowed only if the gas in the star-forming region has a 
metallicity $Z\gsim Z_{cr}=10^{-4}\Zsun$ adopted in this case.

\subsection{Changing the critical metallicity}

\begin{figure*}
  \centerline{\psfig{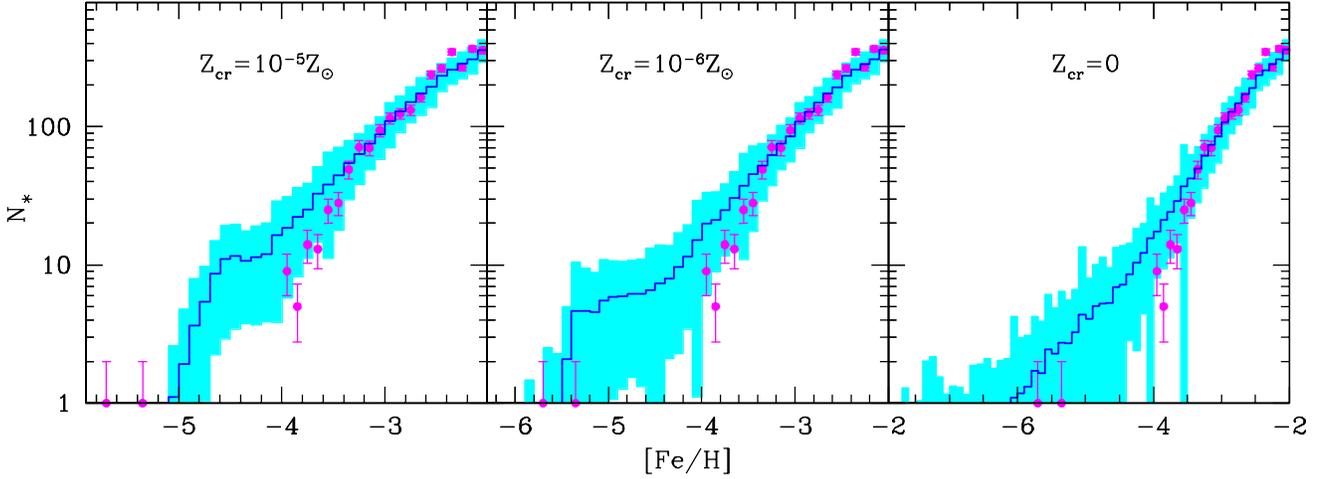}}
  \caption{Same as left panel of Fig.~\ref{fig:6} but for values of $Z_{cr}=10^{-5}, 10^{-6}, 0~ Z_{\odot}$} 
\label{fig:7}
\end{figure*}

In order to bracket the current uncertainty on the determination of $Z_{cr}$, we have
explored the sensitivity of our results to values of $Z_{cr}$ different from the reference 
$Z_{cr}=10^{-4} Z_\odot$ (Fig.~\ref{fig:7}). Independently of the adopted $Z_{cr}$ value, 
the observed global properties of
the MW are reproduced with the same accuracy as in the case $Z_{cr}=10^{-4}Z_{\odot}$. 
We stress again that the choice $Z_{cr}=0$ is equivalent to state that stars 
(including metal-free ones)
form at all times according to a Larson IMF (see Section 3).

The first striking result is that the MDF cutoff is shifted at gradually lower values of 
[Fe/H] when $Z_{cr}$ decreases (for reasons discussed in Sec. 5.1); 
the high-[Fe/H] limit of the distribution appears unchanged, showing, in all cases, 
a very good agreement with the data for [Fe/H]$\geq-3.2$. 
However, the tendency of the data to drop faster than than the model at lower [Fe/H], 
already noticed for $Z_{cr}=10^{-4}\Zsun$, persists. 
In particular, when $Z_{cr}=10^{-5}\Zsun$ the MDF shows a steep increase in 
$-5<$[Fe/H]$<-4.6$ followed by a flat plateau extending up to [Fe/H]$=-4.1$,  
a markedly different shape with respect to the previously analyzed case $Z_{cr}=10^{-4} Z_\odot$. 
This evolution bears the imprint of the GM chemical evolution and to understand
it in detail it is useful to recall the four enrichment phases identified in Fig. \ref{fig:4}. 
During phases in which the GM metallicity grows slowly (as in phase b), stars with approximately equal
[Fe/H] are formed; instead, when the GM metallicity changes rapidly (phases a and c), stars born at slightly 
different cosmological times from accreted gas may have quite different [Fe/H], giving rise to the 
MDF plateau. 
In general, though, the number of stars formed during phase (a) is too small to be appreciated from the
normalized MDF. As $Z_{cr}$ is decreased from $10^{-4} Z_\odot$ to zero, 
phase (b) contracts in time, being almost suppressed for $Z_{cr}=0$.
The correspondingly lower GM metallicity, explains the shift of the low-[Fe/H] MDF 
cutoff to lower values. The metallicity increase during phase (c) brings [Fe/H] to values 
around -3, where Pop~II always dominate the enrichment, independently of the value of $Z_{cr}$.
This physical picture holds qualitatively also for the lower $Z_{cr}$ case discussed below.       
This result can hardly be overlooked, as it shows that the MDF embeds the features imprinted by 
mechanical feedback and reflects the heavy element enrichment history of the MW environment.  

As already mentioned, no stars have been detected so far by the various surveys in the range 
$-5.3< {\rm [Fe/H]} < -4$, a feature that we have dubbed as the ``metallicity desert". 
Not only the model with $Z_{cr}=10^{-5}\Zsun$ does not show such feature, but it also cannot 
account for the presence of the two HMP stars. The latter objects can instead be reproduced 
if $Z_{cr}=10^{-6}\Zsun$. For the reasons explained above, the MDF plateau can extend now 
through the broad range $-5.6\leq$[Fe/H]$\leq -4.2$.
Even in this favorable case, though, the metallicity desert cannot be reproduced. 
If the desert is real or is a mirage produced by selection effects, it is impossible 
to tell at this stage. What is clear though is that hierarchical models like the present 
one have serious difficulties to prevent the formation of stars in the 
desert [Fe/H] range. We will return to this in the discussion.   

Finally, the case $Z_{cr}=0$ is somewhat peculiar. Although it predicts stars for [Fe/H]$<-5.4$, 
as expected from the previous arguments, its MDF shape is different with respect to cases with 
$Z_{cr}\neq 0$, showing a monotonic increase in the range $-6\leq$[Fe/H]$\leq -2$.  
When $Z_{cr}=0$, in fact, no Pop~III/II IMF transition occurs and, as a consequence, the GM 
enrichment proceeds more gradually. Even in this case, though, a considerable number of stars, 
is predicted in the metallicity desert.

\subsection{The parameter $F_{0}$}

The useful parameter, $F_{0}$, has been originally introduced by Oey
(2003) and later improved by Tumlinson (2006). $F_{0}$ is defined as
the number of metal-free stars divided by the total number of stars with [Fe/H]$<-2.5$. 
This parameter allows to quantify the implications of the persisting non-detection of metal-free stars
in the Galactic halo. The observational limit on the value of $F_{0}$
is given by the inverse of the total number of stars with
[Fe/H]$<-2.5$ observed into the galactic halo (where the numerator of
the ratio is fixed equal to 1 to give a non-vanishing limit). Using
the preliminary data by Beers and Christlieb (2006) we find that
$F_{0}^{obs}=1/1152\sim 8.7\times 10^{-4}$.  In Table 2 we present the
values of $F_{0}$ obtained using the models $Z_{cr}=10^{-4}\Zsun$,
$Z_{cr}=10^{-6}\Zsun$, $Z_{cr}=0$.  
\begin{table}\label{table2}
  \begin{center}
    \begin{tabular}{|c|c|}\hline
      \hline 
      $Z_{cr}/\Zsun$ & $F_0$ \\
      \hline 
      \hline
      $10^{-4}\Zsun$ & $6.0\times 10^{-9}$\\
      \hline
      $10^{-6}\Zsun$ & $5.8\times 10^{-9}$\\
      \hline
      $0$ & $7.5\times 10^{-3}$\\
      \hline
    \end{tabular}
  \end{center}
\caption{Values of the $F_0$ parameter (see text) as a function of the assumed $Z_{cr}$.}
\end{table}
Models $Z_{cr} \ge 10^{-6}\Zsun$ give a value of $F_{0}$ compatible with the
observational limit whereas for $Z_{cr}=0$  we find  $F_{0}=7.5\times 10^{-3} >F_{0}^{obs}$. 
In this case, in fact, the first burst of Pop~III stars produces a huge amount of
metal-free, long-lived stars which dramatically pushes $F_{0}$ beyond the allowed range. 
Thus, this parameter allows to robustly rule out models in which the critical metallicity is zero, 
and/or put an important constraint on primordial IMF. In the second option, in order not to conflict   
with the experimental limit, we must set a lower limit on the mass of the first stars equal to
$m_{PopIII}>m_1(z\sim 20)\sim 0.9M_{\odot}$, \ie the stars have a mass larger than the turn-off mass 
at $z\sim 20$, the epoch at which the first star formation events occur in our model.

\subsection{Second generation stars}

\begin{figure*}
  \centerline{\psfig{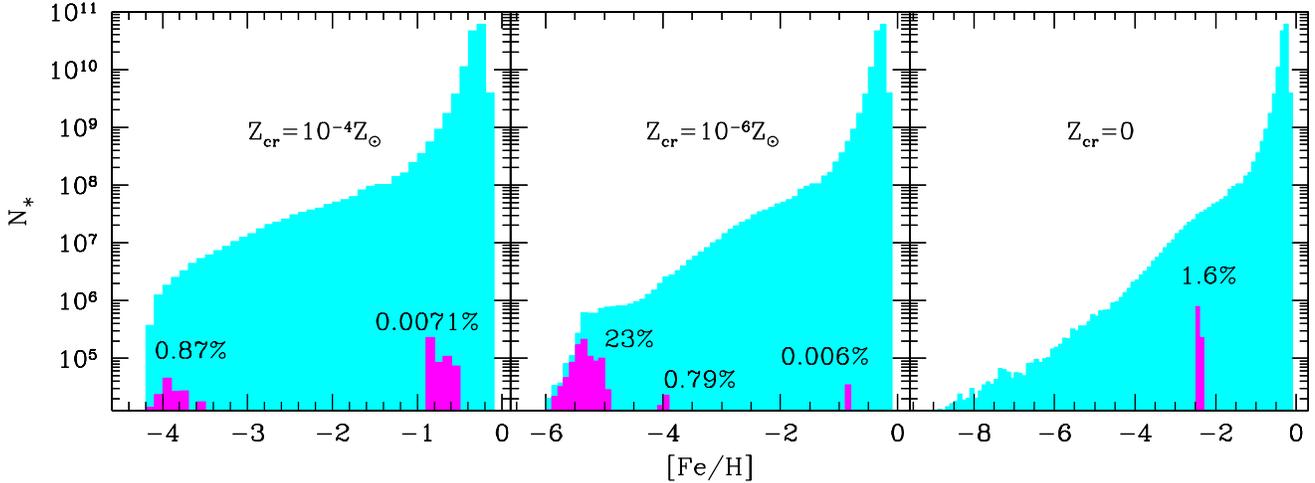}}
  \caption{Impact of second generation stars for the fiducial feedback model
  $Z_{cr}=10^{-4},10^{-6}, 0~ Z_\odot$. In each panel, the highest histogram show       
  the total MDF; the smaller histograms represent the MDFs for the $2^{\rm nd}$-generation 
  stars. The percentages show the fraction of $2^{\rm nd}$-generation stars 
  with respect to the total in the ranges in which they appear.} 
\label{fig:8}
\end{figure*}

An additional information that can be extracted from our study 
is the fraction of ``second generation" (2G) stars, defined as the stars which have been
enriched {\it only} through the nucleosynthetic products of Pop~III stars.
From an observational point of view it is crucial to identify 2G stars, because, by 
studying their elemental composition we could trace back the properties of the
first stars that polluted them.   
 
In Fig.~\ref{fig:8} we identify the MDF regions populated by 2G stars and their fractional number
for three values of $Z_{cr}$. For $Z_{cr}\ge 10^{-6}Z_{\odot}$ 2G stars populate both the lowest 
and the highest [Fe/H] tails of the MDF, \ie they formed both in the accretion and self-enrichment modes. 
If $Z_{cr}=0$, on the contrary, they can be produced only through self-enrichment. 
This is because in this case the first (Type II) SN are not energetic enough to expel 
their metals which remain confined within the parent halos. 

More specifically, the dominant formation channel of 2G stars is accretion when $Z_{cr}=10^{-6}\Zsun$ 
and self-enrichment when $Z_{cr}=10^{-4}\Zsun$, reflecting the different Pop~III/II transition redshift. 
In the case $Z_{cr}=10^{-6}\Zsun$ the transition occurs rapidly as a result of the lower amount
of metals necessary to reach the critical threshold. Hence, 2G stars can be formed through self-enrichment 
only in the few halos hosting Pop~III stars; however, in unpolluted halos accreting gas from the GM, 
2G stars can continue to form until the time at which feedback from Pop~II stars raises the GM metallicity 
substantially.  For the same reasons, the trend is reversed when $Z_{cr}=10^{-4}\Zsun$.

Along with their location on the [Fe/H] axis, the percentages shown in Fig.~\ref{fig:8} quantify
the relative number of 2G stars with respect to the total number of stars in the iron-abundance 
ranges in which 2G stars are found. 
Two points are worth noting. First, the number of 2G stars formed by self-enrichment is negligible; even 
in the case $Z_{cr}=0$, their relative abundance is only $1.6\%$ of the total population of stars with
$-2.5\leq$[Fe/H]$\leq -2.3$; this means that only 10 of the 612 halo stars observed by Beers \& Christlieb 
(2006) in the above [Fe/H] range could be truly 2G stars. Second, it is only in the metallicity bins close 
to the critical metallicity value that the fraction of 2G stars increases substantially, thanks to the 
accretion mode of formation operating at those metallicities. 
However, given the very low number of extremely metal-poor stars available, their absolute number is 
likely to be close to zero. In practice, given the Beers \& Christlieb (2006) sample, out of the 
1150 stars observed in the range $-4\leq$[Fe/H]$<-2.5$, from the results of our model, 
quantified in Table~3, we conclude that, independently on the $Z_{cr}$ value, 
the probability to detect 2G stars is virtually zero.  

\begin{table}\label{table3}
  \begin{center}
    \begin{tabular}{|c|c|}\hline
      \hline 
      $Z_{cr}/\Zsun$ & Number of 2G stars \\
      \hline 
      \hline
      $10^{-4}$ & $1.3$\\
      \hline
      $10^{-6}$ & $0.3$ \\
      \hline
      $0$ & $6\times 10^{-2}$ \\
      \hline
    \end{tabular}
  \end{center}
\caption{Number of 2G stars expected in the Beers \& Christlieb (2006) sample as a function of the 
assumed $Z_{cr}$.}
\end{table}

\section{Summary and Discussion}

Using the newly developed Monte Carlo code, GAMETE, we have followed the gradual build-up of the stellar populations 
and metal enrichment of the Milky Way along its past hierarchical evolution within the concordance 
$\Lambda$CDM model. Adopting simple but physically motivated prescriptions for star formation and 
chemical/mechanical feedback we have first calibrated our model to the observed global properties of 
the MW. Once the model free parameters (the star formation, $\epsilon_{\star}$, 
and the wind, $\epsilon_w$, efficiencies) have been fixed through such calibration, 
the results have been compared against the most recent experimental data 
(Beers \& Christlieb 2006) on the halo stars Metallicity Distribution Function.
We have also explored different values of the critical metallicity, $Z_{cr}$, 
which governs the transition from the Pop~III to Pop~II star 
formation mode (e.g. Schneider \etal 2002).  
The main results are the following:

\begin{enumerate}
\item The global properties of the MW are well reproduced both using a
  closed-box model ($\epsilon_{\star}=0.5$, $\epsilon_{w}=0$, \ie no winds) and a
  model including mechanical feedback ($\epsilon_{*}=0.7$, $\epsilon_{w}=0.2$);
  however, winds are required to dilute metals in the Galactic Medium and hence 
  reproduce the wide range of [Fe/H] values characterizing halo stars (see Fig. 3).
\item Stars with [Fe/H]$<-2.5$ form in halos accreting GM gas enriched by earlier supernova explosions.
\item The fiducial (feedback included) model for which $Z_{cr} =
  10^{-4} Z_\odot$, $m_{PopIII}=200\Msun$ 
  provides a very good fit to the MDF, including the cutoff observed at [Fe/H]$=-4$. 
  However: the existence of the two HMP stars with [Fe/H]$<-5$ cannot be explained within this model.
\item Models with $Z_{cr}\le 10^{-6}\Zsun$ can account for the two HMP stars at the price of
  overpopulating the ``metallicity desert", \ie the range $-5.3<$ [Fe/H] $<-4$ in which no 
  stars have been detected so far.  
\item The current non-detection of metal-free stars, giving a value of the parameter $F_{0}=1/1152$, 
  robustly constrains either $Z_{cr}$ to be larger than zero or the masses of the first stars
  to be $m_{PopIII}>0.9\Msun$.
\item The statistical impact of truly second generation stars, \ie stars forming out of gas 
  polluted {\it only} by metal-free stars, is negligible. For example, in the Beers \& Christlieb 
  (2006) sample only (at best) 1-2 stars could retain such metal-free nucleosynthetic imprint.
\item Independently of $Z_{cr}$, $60\%$ of metals in the Galactic Medium are ejected through winds 
 by halos with masses $M<6\times 10^9\Msun$, thus showing that low-mass halos are the dominant 
 population contributing to cosmic metal enrichment.    
 \end{enumerate}

The previous results provide a self-consistent and coherent physical scenario for the 
formation and evolution of the Galaxy in a cosmological context. The model also explains
in detail how the transition to Pop~II stars occurred and clarifies by what mechanisms (mostly
self-enrichment and gas accretion) the various regions of the MDF have been populated.

In spite of these successes not all is clear and additional work will be necessary to 
understand some puzzling issues.  No doubt that the most outstanding one is the unexplained
existence of the ``metallicity desert". The problem in its simplest form can be stated as 
follows: in order to produce low-mass stars with [Fe/H] $<-5$, the value of $Z_{cr}$ must be 
decreased to similar levels (or below); as a consequence, an uncomfortably large number
of stars is predicted by such a model in the desert.  Assuming that the extremely low [Fe/H] of the 
two stars (which are very CNO-rich and have a total $Z\gg Z_{cr}$) is not a result of some peculiar 
nucleosynthetic history, ways to quench the formation of stars in the desert region must be devised.
If the Pop~III star formation episodes indeed spread their metals into the GM to reach a metallicity
levels of ${\rm [Fe/H]} \approx -5$, they could also heat the gas to about $T_h \gsim 10^5$~K. In such  
preheating scenario (Madau, Ferrara \& Rees 2001) gas accretion on virializing halos with $T_{vir}< T_h$
would be prevented, whereas star formation and self-enrichment could continue in already collapsed
structures, which would increase the GM metallicity through their winds. It is only when dark matter 
halos of sufficiently large mass become numerous that accretion onto them could start again, now producing
stars with [Fe/H]$> -4$. Translated in redshift, and guided by Fig. 4, the metallicity desert could
translate in a gap of $\Delta z \approx 5$ at $z>10$ in which accretion is halted by the preheating,
another aspect of the mechanical feedback, which once again is found to play a crucial role in the problem.   

The Holy Graal of the various halo star surveys has undoubtedly been the detection of a truly metal-free
star. As these objects have so far escaped detection, growing attention has been posed in the recent
years on probing the nucleosynthetic patterns of extremely metal-poor stars, with the hope that these
objects are born from gas polluted purely by (a single ?) metal-free SN. Unfortunately, our study shows
that these second generation stars are no less elusive than their predecessors, having number frequencies
that are so low to be zero even in the largest available sample used in the present analysis. 
These findings open a different perspective on the recent data by Cayrel \etal 2004 (later interpreted
by Chiappini \etal 2005). Using a sample of 30 stars in the range $-4.1<$ [Fe/H] $< 1.7$ they measured
the abundances of 17 elements from C to Zn, finding that these stars have an extremely low abundance ratio
scattering. This is obviously contrary to what expected if these stars are direct descendants of the 
first generation(s) of stars formed after the Big Bang. Our results indicate that the stars used for the
analysis are almost certainly {\it not} second generation stars; instead, they represent a more mature 
population in which the abundance scatter has been smoothed out by several re-processing cycles of the gas.

A number of assumptions and approximations have also been done, mainly dictated by our persisting ignorance
of some physical processes. These also need to be improved in future
studies. The most critical one is probably the perfect mixing
approximation which may acclerate the extinction of metal free stars
with respect to the epoch $z=11$ predicted by our results. This
problem is partially alleviated by the spread induced by the
stochastic nature of merger histories which appears to be similar to
that found by sophisticated numerical simulations of mixing in
individual galaxies (Mori \& Umemura, 2006). An additional key
hypothesis concerns the IMF of Pop~III stars which has been taken to
be either a $\delta$-function in the SN$_{\gamma\gamma}$ mass range
$140\Msun<m_{PopIII}<260\Msun$ or a standard Larson IMF when
$Z_{cr}=0$.  Given the strong [Fe/H] dependence of $m_{PopIII}$ we
have assumed as reference value $m_{PopIII}=200\Msun$ but we have also
explored the implications of adopting the two extreme values of
$140\Msun$ and $260\Msun$ in the Appendix.
A Gaussian Pop~III IMF centered in $200\Msun$, as suggested by some theoretical
works (see Nakamura \& Umemura 2001; Scannapieco, Schneider \& Ferrara
2003) might help explaining the concomitant existence of the two HMP stars
and the metallicity desert.

Additional hypothesis have been made. Star formation has been take to occur in burst only,
rather than in a continuous manner. This assumption is good for low-mass halos, but
might become increasingly less accurate in larger galaxies. However, the fact that the
global properties of the MW are well reproduced, seems to indicate that this is not
a major concern.  Also, instantaneous recycling has been used. This is definitely
a fair approximation for very short-lived Pop~III stars, but might be a relatively poor
description of the chemical evolution driven by more evolved generations.   
This might induce an overestimate of the metallicity, and consequently, an earlier
Pop~III/II transition than the actual one; when averaged over the present Hubble time,
though, this problem should not affect the derived MDF properties. 

We finally like to comment on the differences between our results and those obtained by Tumlinson (2006).
The treatment of chemical and mechanical feedback is probably more physical in our scheme,
allowing metals/matter transfer between halos and the GM. Within our model, we find that the  
Pop~III/II transition is a relatively sudden event, rather than a smooth one extending
down to $z<6$ as Tumlinson found. This might be the result of the perfect mixing
approximation that we have used, whose broad validity is supported by the afore-mentioned
analysis of the metal filling factor evolution. Another difference between the two models 
resides in the calibration method: we select the best models in terms of
the two free parameters $\epsilon_{\star}$ and $\epsilon_{w}$ as those reproducing 
{\it simultaneously} all the observed global properties of the MW, rather than 
fixing them separately. As an indicator of the gas metallicity, we have used observations
of gas in the Local Group (as for example the High Velocity Clouds), rather than the
general IGM which is very likely not representative of the Galactic environment we are investigating. 

In spite of these differences, there are no major tensions between the results of the two studies.
It is worth noting that Tumlinson also identifies $Z_{cr}\approx 10^{-4}\Zsun$ as the most 
probable transition threshold, although his analysis could not discriminate between largely 
different $Z_{cr}$ values, given the poorer sample statistics; also, the two HMP stars are not 
taken into account in his analysis. 

\section*{Acknowledgements}
We are grateful to T. Beers and N. Christlieb for providing us with their MDF data in 
advance of publication, and the anonymous referee for his/her
insightful comments. We thank P. Bonifacio, N. Christlieb, A. Frebel,
A. Helmi, N. Karlsson, E. Scannapieco, J. Tumlinson and the
DAVID\footnote{{\tt
    www.arcetri.astro.it/science/cosmology/index.html}}  
members for enlightening discussions. The work has been completed
during the program (INT-06-2a) of the Institute for Nuclear Theory,
Seattle, whose support is gratefully acknowledged.
\bibliographystyle{mn}
\bibliography{biblio}

\appendix
\section{Varying the Pop~III mass}
The models in Sec.~6 assume $m_{PopIII}=200\Msun$ and $Z_{cr}>0$; we
now explore the effects of setting $m_{PopIII}$ equal to
$140,260\Msun$, respectively. These two values are the extremes of the
SN$_{\gamma\gamma}$ mass range and correspond to the minimum and
maximum iron yield of $Y_{Fe}=2.8\times 10^{-15}$ and $Y_{Fe}=0.45$. 
We recall instead that $Y=0.45$, independently of $m_{PopIII}$.

Following the same arguments using in Sec.~5.1 we have checked that
models without feedback largely fail to reproduce the overall MDF
shape also for $m_{PopIII}=140, 260\Msun$; then need for feedback is
independent of the adopted Pop~III stellar mass. We have derived the
MDF for galactic halo stars for these two cases; this is shown in
Fig.~9 for the usual values of the two free parameters
$\epsilon_*=0.7$, $\epsilon_w=0.2$. 

For $m_{PopIII}=140\Msun$, the shape of the MDFs is completely
independent of $Z_{cr}$ and closely resembles the behavior found 
for the case $Z_{cr}=0$. This result from the extremely low $Y_{Fe}$
of $140\Msun$ stars which makes their contribution negligible to that
of Pop~II stars which are then controlling the MDF shape.

If $m_{PopIII}=260\Msun$ instead, the MDF cutoff are quite sensitive
to the value of $Z_{cr}$, shifting to towards lower [Fe/H] values with
decreasing $Z_{cr}$. Such trend is very similar to the one already found 
for models with $m_{PopIII}=200\Msun$. 
However in this case, given the higher $Y_{Fe}$,
the cutoff is located at higher [Fe/H]. For this reason the observed
data are better reproduced by model with $Z_{cr}\leq 10^{-5}\Zsun$. 
Note however that even these models cannot account for the presence of
the two HMP stars. 

In conclusion the MDF is sensitive to the value of $m_{PopIII}$.
However, as long as $m_{PopIII}$ is within the SN$_{\gamma\gamma}$
mass range, it is always possible to find satisfactory fits to the
data by suitable changing $Z_{cr}$ within its uncertainty interval 
$Z_{cr}=10^{-5\pm1}\Zsun$.

Most importantly both the value of $F_0$ (Sec.~6.2) and the number of
2G stars (Sec.~6.3) are practically independent of the value of 
$m_{PopIII}$. In particular, using the two best models
$m_{PopIII}=140\Msun$, $Z_{cr}=10^{-4}\Zsun$ and
$m_{PopIII}=260\Msun$, $Z_{cr}=10^{-6}\Zsun$ we obtain 
$F_0=6.88\times 10^{-9}$ and $6.3\times 10^{-9}$, respectively. 
The expected number of 2G stars in the Beers \& Christlieb (2006)
sample, to be compared with Tab.~3, is equal to $3.45\times 10^{-2}$
and $2.2\times 10^{-2}$, respectively.     

\begin{figure*}
  \centerline{\psfig{figure=FIG_9.epsi,width=13.1cm,angle=0}}
  \caption{Same as left panel of Fig.~6 but for different values of
  $m_{PopIII}=140, 260 \Msun$ and $Z_{cr}=10^{-4},10^{-5},10^{-6}\Zsun$.}
\label{fig:9}
\end{figure*}

\label{lastpage}

\end{document}